\documentclass[%
reprint,
amsmath,amssymb,
aps,
floatfix,
]{revtex4-2}
\usepackage{graphicx}
\usepackage{dcolumn}
\usepackage{bm}
\usepackage{subcaption}
\begin{document}

\begin{abstract}
The Laser Interferometer Space Antenna (LISA) will be a space-borne gravitational wave (GW) detector to be launched in the next decade. Central to LISA data analysis is time-delay interferometry (TDI), a numerical procedure which drastically reduces otherwise overwhelming laser frequency noise. LISA data analysis is usually performed on sets of TDI variables, e.g. Michelson variables $(X, Y, Z)$ or quasiorthogonal variables $(A, E, T)$. We investigate a less standard TDI variable denoted $\kappa$ which depends on time, or frequency, and two parameters $(\beta, \lambda)$. This so-called coronagraphic TDI variable has the singular property of canceling GW signal when $(\beta, \lambda)$ tend to the sky position of the GW source. Thanks to this property, coronagraphic TDI has the potential to be an efficient model-agnostic method for sky localization of GW sources with LISA. Those characteristics make it relevant for low-latency searches and a possible glitch veto. Although briefly discussed in the literature, coronagraphic TDI has only been tested on theoretical grounds. In this paper we validate the applicability of $\kappa$ to sky localization of typical LISA sources, namely Galactic binaries (GBs) and massive black hole binaries (MBHBs), when considering a simplified LISA instrument. The goal of this paper is to pave the way for applications of coronagraphic TDI to practical LISA data analysis problems.
\end{abstract}

\title{Coronagraphic time-delay interferometry:\\ characterization and updated geometric properties}%
\author{R. Costa Barroso}%
\email{raissa.costa-barroso@unicaen.fr}%
\author{Y. Lemière}%
\author{F. Mauger}%
\affiliation{%
Université de Caen Normandie, ENSICAEN, CNRS/IN2P3,\\
LPC Caen UMR6534, F-14000 Caen, France.
}%
\author{Q. Baghi}%
\affiliation{%
Université Paris Cité, CNRS, Astroparticule et Cosmologie, Paris F-75013, France.
}%
\date{\today}

\maketitle

\section{Introduction}
The Laser Interferometer Space Antenna (LISA) is a space-borne gravitational wave (GW) detector to be launched in the next decade. LISA is a constellation of three spacecraft forming an equilateral triangle orbiting around the Sun and trailing the Earth. With an armlength of $2.5 \cdot 10^6$ km, LISA is designed to be sensitive to GWs in the mHz band and is expected to observe an unprecedented number of sources: quasi-monochromatic sources as Galactic binaries (GBs), transient sources such as massive black hole binaries (\mbox{MBHBs}), extreme mass ratio inspirals, stochastic GW backgrounds and potentially unmodeled sources \cite{LISA}. The wealth of physics carried by LISA data comes with challenging data analysis problems.

Central to LISA data analysis is time-delay interferometry (TDI), an on-ground processing technique which recombines multiple interferometric on-board measurements and drastically reduces otherwise overwhelming laser frequency noise. This algorithm gives rise to multiple TDI variables which have been discussed across the LISA literature (see for instance \cite{PhysRevD.105.062006} and references therein). Notably, it has been shown that so-called first generation TDI variables, that is when neglecting the relative motion of one spacecraft with respect to the other, form a module: the module of syzygies \cite{PhysRevD.65.102002}. As a consequence, any TDI variable in the module can be expressed as a combination of four generators, for instance the Sagnac generators  $(\alpha, \beta, \gamma, \zeta)$. This raises the question: is there an optimal combination?

This question has been made more precise and answered in \cite{PhysRevD.66.122002, PhysRevD.70.062002, K_Rajesh_Nayak_2003, PhysRevD.68.122001}. In particular, one can be interested in the TDI combination which minimizes signal to noise ratio (SNR) for a given direction in the sky. This TDI variable, denoted by $\kappa$, depends on time and on two parameters $(\beta, \lambda)$ which are interpreted as latitude and longitude. In the idealized noiseless scenario of a single localized GW source, when $(\beta, \lambda)$ approach the position of the GW source in the sky $(\beta_\star, \lambda_\star)$ this TDI variable tends to zero. This property implies that $\kappa$ behaves very much like a coronagraph, hence the name coronagraphic TDI \cite{JYV}.  

From a data analysis perspective, this means one can scan the sky seeking for the point where coronagraphic TDI tends to zero and, in principle, estimate the position of the source of the incoming GW. In contrast to standard GW data analysis methods which rely on waveform template generation, here estimation of sky position parameters is performed without making any assumption on the nature of the source. Because sky localization is a crucial piece of information to be communicated to astronomical observatories when aiming at multi-messenger detection, developing a fast model-agnostic method for estimating the position of a GW source is relevant for low-latency searches. Another possible application is to use $\kappa$ as a glitch veto.

The goal of this paper is to assess the applicability of $\kappa$ to data analysis problems, in particular to sky localization of GW sources. We focus on evaluating the method's sensitivity to both source parameters and to the evolution of the geometry of the LISA constellation. It foresees the design of a realistic low-latency algorithm based on $\kappa$, which is out of the scope of this study. This paper is organized as follows. In Sec. \ref{sec:tdi} we briefly present the construction of coronagraphic TDI as in \cite{PhysRevD.70.062002, PhysRevD.68.122001, JYV} and demonstrate the method. In Sec. \ref{sec:signal} we characterize the proposed sky localization technique by estimating, under simplifying assumptions, the angular resolution which can be achieved. In Sec. \ref{sec:detector} we test coronagraphic TDI on an analytic noise model. Finally, we shift to realistic orbits and examine $\kappa$'s response in this scenario.

\section{Coronagraphic TDI} \label{sec:tdi}

\subsection{Principle} \label{sec:definition}
Coronagraphic TDI is essentially a linear combination of standard TDI variables with carefully chosen coefficients. It relies on the fact that the TDI algorithm uses information from different spacecraft in a specific way. As a consequence, a GW coming from a given direction at a given frequency has a unique imprint on a given TDI variable. This characteristic imprint can then be used to construct coefficients which take advantage of the structure imposed by TDI to block the signal. In this section we present the basic steps of the derivation of coronagraphic TDI in frequency domain following references \cite{PhysRevD.70.062002, PhysRevD.68.122001, JYV}. More details can be found in App. \ref{sec:app_definition}.

To begin with, it is important to stress that this derivation relies on the fact that first generation TDI variables form a module and as a consequence any first generation TDI variable is a combination of four generators, such as Sagnac generators $(\alpha, \beta, \gamma, \zeta)$. Coronagraphic TDI is one specific combination of these generators. Moreover, a fundamental assumption to this derivation is that LISA orbits are static. This is of course not true in reality but is a valid approximation for relatively short periods of time. In fact, LISA arm lengths are expected to vary at a rate of up to 12 m/s \cite{redbook}, which corresponds to a variation of 1\% in arm length after a month. Also note that this derivation holds for both equal and unequal arms configurations.

At this point, it is useful to fix some notations and definitions. For instance, $L_{ij}$ is the light-travel time of a photon emitted at spacecraft $j$ and received at spacecraft $i$. Similarly, the unit vector $\hat{n}_{ij}$ points from spacecraft $j$ to spacecraft $i$.

Under the assumptions described above, one can express an arbitrary first generation TDI variable in frequency domain as \cite{PhysRevD.70.062002}:
\begin{equation}
    \tilde{\kappa} = a_\alpha \tilde{\alpha} + a_\beta \tilde{\beta} + a_\gamma \tilde{\gamma} + a_\zeta \tilde{\zeta}
\end{equation}
where $a_\alpha$, $a_\beta$, $a_\gamma$ and $a_\zeta$ are arbitrary complex functions and $(\tilde{\alpha}, \tilde{\beta}, \tilde{\gamma}, \tilde{\zeta})$ are the Fourier transforms of $(\alpha, \beta, \gamma, \zeta)$. In principle, the derivation that follows can be done with the four generators. However, doing so incurs additional mathematical complications which can be avoided if one restricts to $(\alpha, \beta, \gamma)$. In fact, because $\zeta$ is related to the other generators, except for a finite number of frequencies, in frequency domain $\tilde{\zeta}$ is just a combination of $(\tilde{\alpha}, \tilde{\beta}, \tilde{\gamma})$ up to a factor \cite{PhysRevD.66.122002}. Then SNR simply writes \cite{PhysRevD.70.062002}:
\begin{equation}
    \mathrm{SNR}_{\kappa}^2 = \int_{f_l}^{f_u} \frac{|a_\alpha \tilde{\alpha}_{d} + a_\beta \tilde{\beta}_{d} + a_\gamma \tilde{\gamma}_{d}|^2}{\langle |a_\alpha \tilde{\alpha}_{n} + a_\beta \tilde{\beta}_{n} + a_\gamma \tilde{\gamma}_{n}|^2 \rangle}\mathrm{d}f \label{eq:SNR}
\end{equation} 
where the subscript $d$ stands for data and the subscript $n$ for noise. Integration is performed over frequency $f$ within the interval $(f_l,f_u)$ corresponding to the LISA band. The brackets in the denominator denote an ensemble average. Noticing that $\mathrm{SNR}_\kappa^2=0$ if and only if the numerator is zero, the original problem reduces to solving  
\begin{equation}
    \tilde{\kappa}_{d} = a_\alpha \tilde{\alpha}_{d} + a_\beta \tilde{\beta}_{d} + a_\gamma \tilde{\gamma}_{d} = 0. \label{eq:problem}
\end{equation}

For the sake of the argument, consider the noiseless case where data contains the signal of a single localized monochromatic source of frequency $f_\star$ and sky position $(\beta_\star, \lambda_\star)$. In this case, we replace the index $d$ in Eq.\eqref{eq:problem} by the index $gw$. Then one can write $\tilde{\alpha}_{gw}$ as
\begin{equation}
    \tilde{\alpha}_{gw} = \alpha_+ \tilde{h}_+ + \alpha_\times \tilde{h}_\times, \quad \alpha_{+,\times} = \sum_{ij} \alpha_{ij} \xi_{ij}^{+,\times} \label{eq:decomposition}
\end{equation}
with indices $ij$ running through $\{31, 13, 23, 32, 12, 21\}$, that is the arms of the constellation. The $\alpha_{ij}$ carry all TDI information, that is information about time delays, while the $\xi_{ij}^{+, \times}$ denote the antenna pattern functions.  Explicit expressions for $\alpha_{ij}$ and $\xi^{+, \times}_{ij}$ are given in App. \ref{sec:app_definition}. Furthermore, we omit the dependence of $\alpha_{ij}$ and $\xi_{ij}^{+, \times}$ on frequency and sky position. In fact, in Eq.\eqref{eq:decomposition} one should write
\begin{equation}
    \alpha_{ij}(f_\star, \beta_\star, \lambda_\star) \quad \text{and} \quad \xi_{ij}^{+, \times}(\beta_\star, \lambda_\star).
\end{equation}
Notice that the antenna pattern functions do not depend on frequency because LISA orbits are assumed to be static. As a consequence, the $\alpha_{+, \times}$ inherit the same dependence on frequency and sky position. To be complete, one should write
\begin{equation}
    \alpha_{+,\times}(f_\star, \beta_\star, \lambda_\star).
\end{equation}
Decomposing TDI variables $\tilde{\beta}_{gw}$ and $\tilde{\gamma}_{gw}$ in a similar way to Eq.\eqref{eq:decomposition}, one can rewrite Eq.\eqref{eq:problem} as
\begin{equation}
    \begin{split}
        \tilde{\kappa}_{gw} &= (a_\alpha \alpha_+ + a_\beta \beta_+ + a_\gamma \gamma_+) \tilde{h}_+ \\
        & \quad + (a_\alpha \alpha_\times + a_\beta \beta_\times + a_\gamma \gamma_\times) \tilde{h}_\times = 0. \label{eq:expansion}
    \end{split}
\end{equation}

Then introducing the vectors
\begin{equation}
    \vec{A} = \begin{pmatrix}
        a_\alpha \\
        a_\beta \\
        a_\gamma
    \end{pmatrix}
    \quad \text{and} \quad 
    \vec{P}_{+, \times} = \begin{pmatrix}
        \alpha_{+, \times} \\
        \beta_{+, \times} \\
        \gamma_{+, \times}
    \end{pmatrix}
\end{equation}
Eq.\eqref{eq:expansion} becomes
\begin{equation}
    \tilde{\kappa}_{gw} = (\vec{A}\cdot\vec{P}_+)\tilde{h}_+ + (\vec{A}\cdot\vec{P}_\times)\tilde{h}_\times = 0. \label{eq:vector}
\end{equation}
Because Eq.\eqref{eq:vector} should hold for any $\tilde{h}_{+, \times}$, in particular for $\tilde{h}_{+,\times} \neq 0$, the problem stated in Eq.\eqref{eq:problem} amounts to solving
\begin{equation}
    \left\{
    \begin{array}{ll}
        \vec{A} \cdot \vec{P}_+ = 0 \\
        \vec{A} \cdot \vec{P}_\times = 0.
    \end{array}
\right.
\end{equation}
That is $\vec{A}$ is simply the vector product between the two projectors $\vec{P}_{+}$ and $\vec{P}_{\times}$,
\begin{equation}
    \vec{A} = \vec{P}_+ \times \vec{P}_\times = \begin{pmatrix}
        \beta_+ \gamma_\times - \gamma_+ \beta_\times \\
        \gamma_+ \alpha_\times - \alpha_+ \gamma_\times \\
        \alpha_+ \beta_\times - \beta_+ \alpha_\times
    \end{pmatrix}. \label{eq:A}
\end{equation}

By moving away from $(f_\star, \beta_\star, \lambda_\star)$ and evaluating $\vec{A}$ at any $(f, \beta, \lambda)$, one obtains a TDI variable which depends on three parameters $(f, \beta, \lambda)$ and has the property to vanish when $(f, \beta, \lambda) \rightarrow (f_\star, \beta_\star, \lambda_\star)$. This property holds for any monochromatic GW source and can be extended to non-monochromatic sources by considering a set of frequencies $\{f_\star\}$. No additional assumptions on $\tilde{h}_{+,\times}$ are required. This leads to the definition of $\tilde{\kappa}$, also given in Eq.(24) in \cite{PhysRevD.70.062002},
\begin{equation}
    \tilde{\kappa}(f, \beta, \lambda) = \vec{A}(f, \beta, \lambda) \cdot \vec{D}(f). \label{eq:def}
\end{equation}
where $\vec{D}(f) = (\tilde{\alpha}(f), \tilde{\beta}(f), \tilde{\gamma}(f))^T$. Notice that $\vec{A} = \vec{P}_\times \times \vec{P}_+$ is also a solution to Eq.\eqref{eq:problem} so $\tilde{\kappa}$ has two minima, one at $(f_\star, \beta_\star, \lambda_\star)$ and another at $(f_\star, -\beta_\star, \lambda_\star)$. Lastly, as showed in \cite{PhysRevD.70.062002}, by taking the inverse Fourier transform of $\tilde{\kappa}$ one can write $\kappa(t, \beta, \lambda)$ directly in time domain. More details are provided in App. \ref{sec:app_definition}. 

It is worth highlighting the fact that coronagraphic TDI, as defined in Eq.\eqref{eq:def}, is the scalar product between two vectors. One which can be interpreted as a vector pointing at various directions of the sky, vector $\vec{A}$, and another which bears the data, vector $\vec{D}$. The very structure of this paper is motivated by this observation. In Sec. \ref{sec:signal} we examine $\tilde{\kappa}$'s sensitivity to changes in the content of $\vec{D}$, while in Sec. \ref{sec:detector} we are interested in identifying how changes in $\vec{A}$ impact $\tilde{\kappa}$. On a practical note and within the context of low-latency searches, the structure of $\tilde{\kappa}$ implies that $\vec{A}$ can be precomputed independently from $\vec{D}$. Thus greatly improving the efficiency of the method. 

\subsection{Demonstration} \label{sec:demonstration}
Coronagraphic TDI was implemented and tested on a sine wave in \cite{PhysRevD.70.062002}. Here, we reimplement it and interface it with LISA simulation tools recently developed by the LISA Consortium, namely \textsc{LISA Orbits} \cite{orbits}, \textsc{LISA GW Response} \cite{gwresponse} and \textsc{PyTDI} \cite{pytdi}. We test our implementation on the following datasets:
\begin{enumerate}
    \item VGBs simulated with \textsc{LISA GW Response}. We consider the 16 VGBs listed in \cite{kupfer2023lisa}.
    \item MBHBs simulated with the \textsc{LISA Data Challenge} software \cite{LDC} using \texttt{IMRPhenomD} waveforms. The response of LISA arms is simulated with \textsc{LISA GW Response}. We consider the 15 MBHBs from the Sangria (LDC-2a) dataset \cite{sangria}\footnote{In this dataset MBHBs' parameters are randomly drawn from a catalog provided by the LISA Astrophysics Working Group.}.
\end{enumerate}
Moreover, in our simulations we assume a static constellation with unequal arm lengths. Importantly, we suppose the constellation to be lying on the ecliptic plane. In fact, with coronagraphic TDI the position of the source is obtained in the LISA frame but in this case the latter and the ecliptic plane overlap. The assumption on the position of the LISA constellation is discussed in more detail in Sec. \ref{sec:detector}. All simulations are noiseless, contain only one signal and last for 5.8 days.

For each dataset, we test coronagraphic TDI in time domain and frequency domain. In particular, in frequency domain we produce a sky map as seen by $\tilde{\kappa}$. The figures described below give a better understanding of how coronagraphic TDI works in practice. They are also the starting point of a sky localization method based on coronagraphic TDI.

\begin{figure}[ht!]
    \begin{subfigure}[b]{0.48\textwidth}
        \centering
        \includegraphics[scale=0.28]{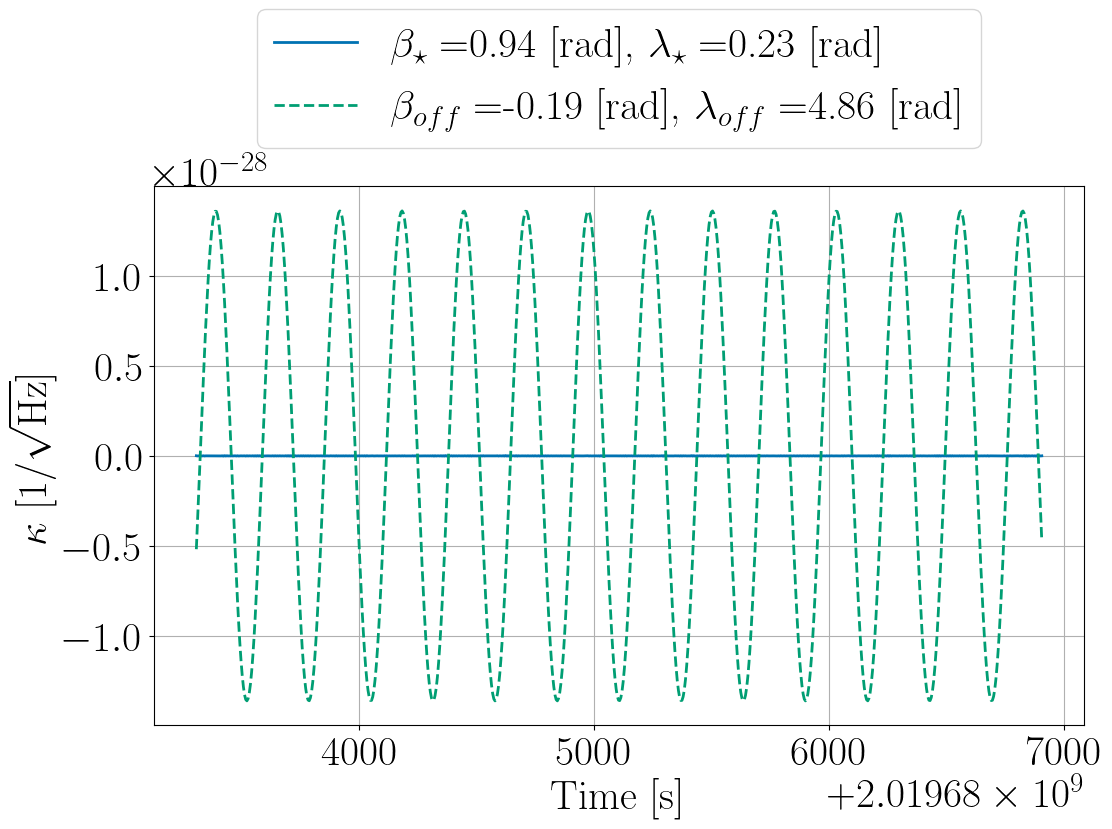}
        \caption{Coronagraphic TDI in time domain for VGB ZTFJ2243.}
        \label{fig:time_vgb}
    \end{subfigure}
    \hfill
    \begin{subfigure}[b]{0.48\textwidth}
        \centering
        \includegraphics[scale=0.28]{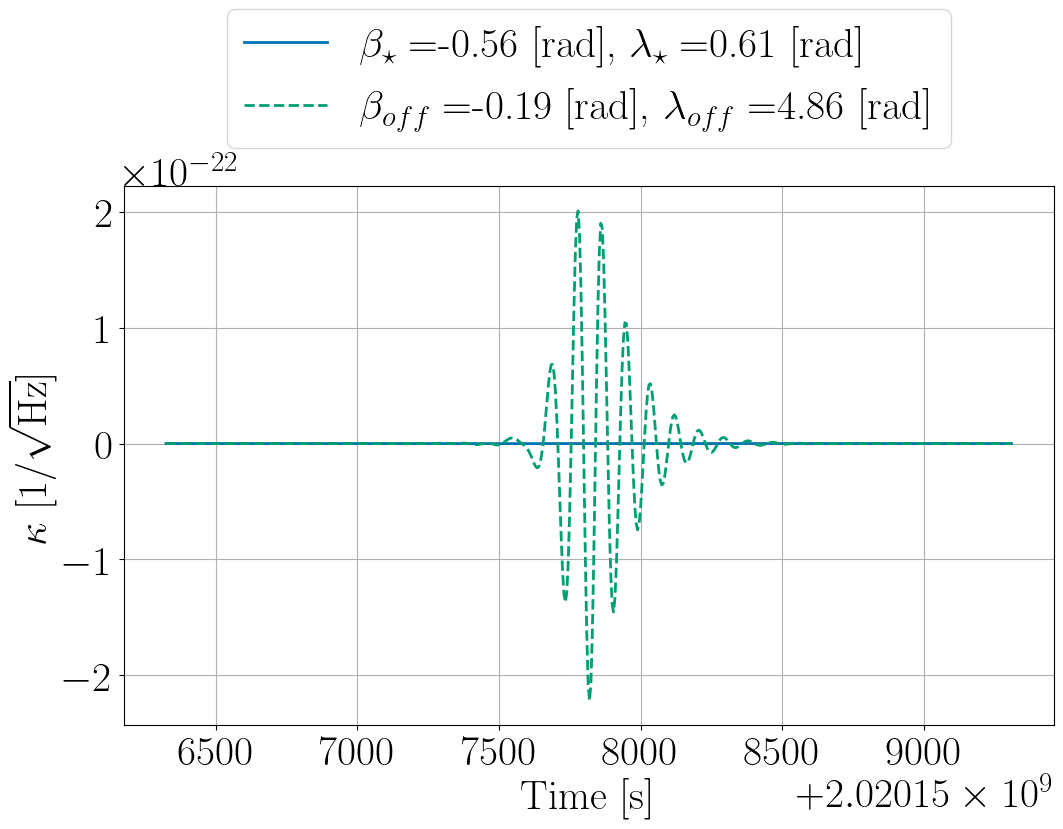}
        \caption{Coronagraphic TDI in time domain for MBHB\_00.}
        \label{fig:time_mbhb}
    \end{subfigure}
    \caption{Coronagraphic TDI in time domain.}
    \label{fig:proof_time}
\end{figure}

In Fig. \ref{fig:proof_time}, we show two examples of $\kappa$ computed in time domain, one for a VGB and another for a MBHB. In each example, we validate the suppression of the signal by comparing the values of $\kappa$ when pointing at two directions: the position of the source and some random direction $(\beta_{off}, \lambda_{off})$ different from the position of the source. In both figures, the signal in $\kappa(t, \beta_\star, \lambda_\star)$ is suppressed relative to $\kappa(t, \beta_{off}, \lambda_{off})$, .

\begin{figure}[ht!]
    \begin{subfigure}[b]{0.48\textwidth}
        \centering
        \includegraphics[scale=0.28]{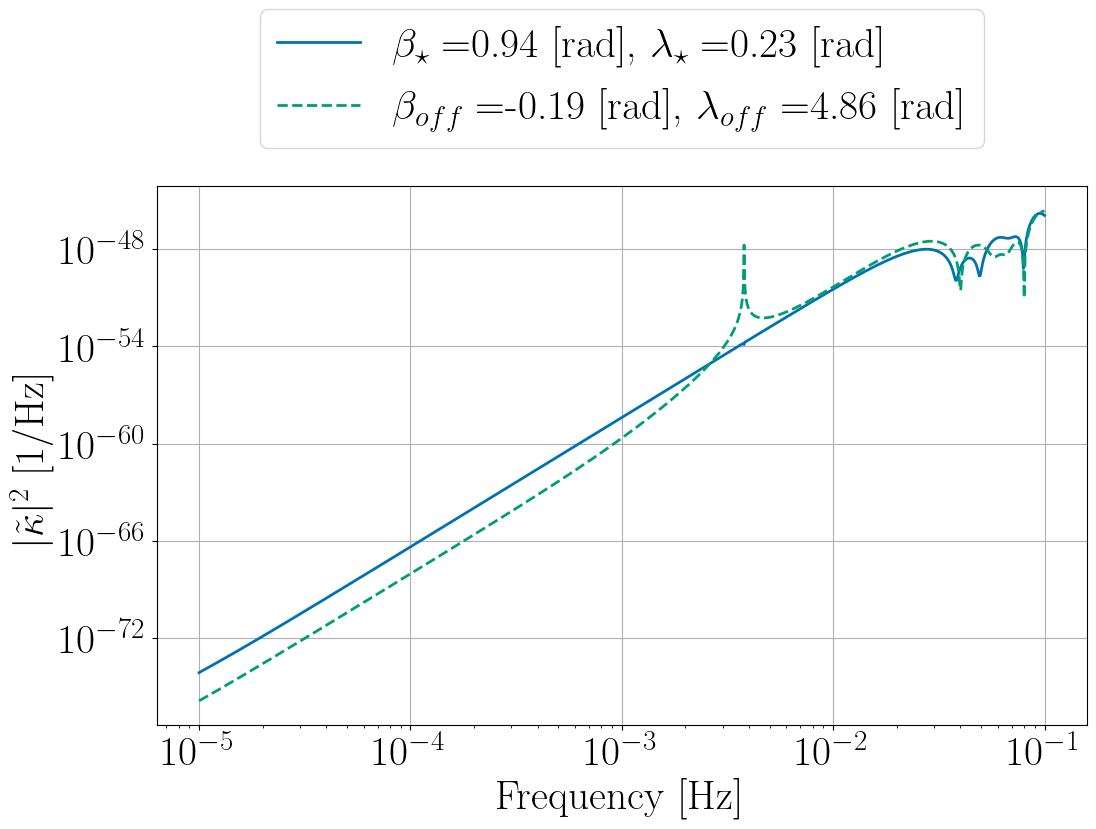}
        \caption{Coronagraphic TDI in frequency domain for VGB ZTFJ2243.}
        \label{fig:freq_vgb}
    \end{subfigure}
    \hfill
    \begin{subfigure}[b]{0.48\textwidth}
        \centering
        \includegraphics[scale=0.28]{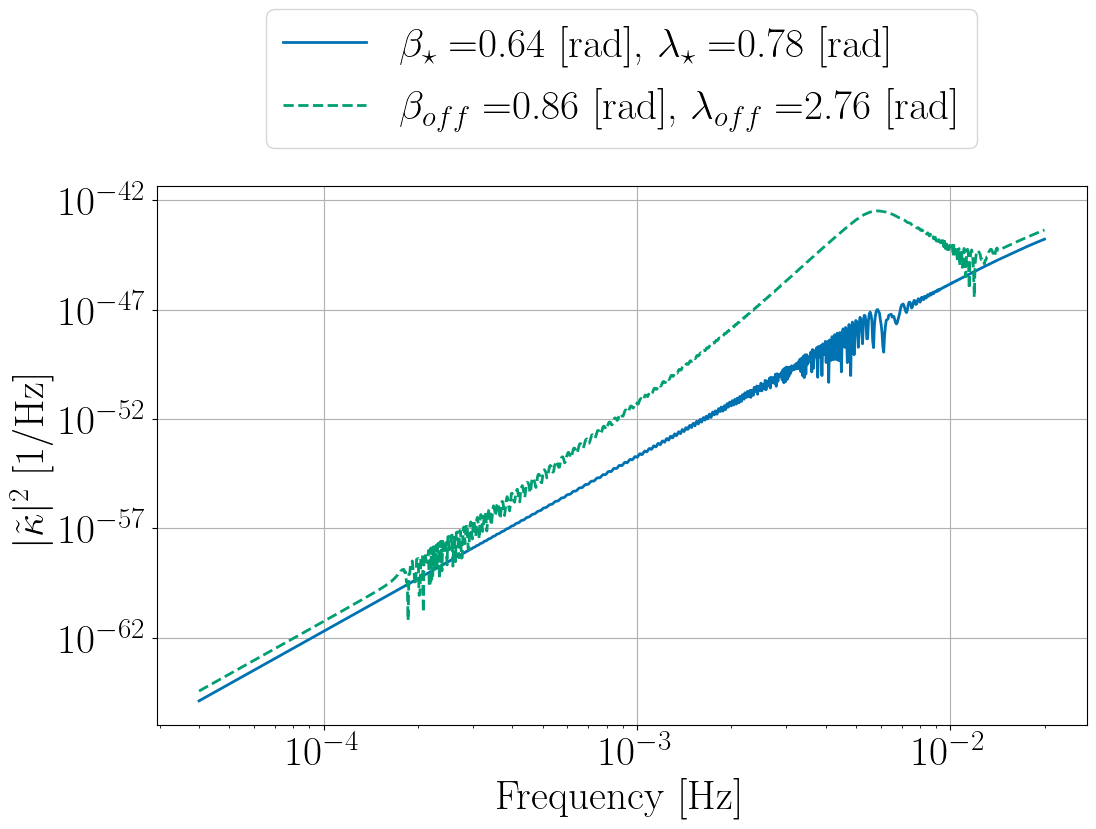}
        \caption{Coronagraphic TDI in frequency domain for MBHB\_00.}
        \label{fig:freq_mbhb}
    \end{subfigure}
    \caption{Coronagraphic TDI in frequency domain.}
    \label{fig:proof_freq}
\end{figure}

Likewise, Fig. \ref{fig:proof_freq} shows results in frequency domain. In the quasi-monochromatic case the signal is characterized by a frequency $f_0$. For VGB ZTFJ2243 this frequency is at $f_0 = 3.788$ mHz \cite{kupfer2023lisa}. In Fig. \ref{fig:freq_vgb} we observe that around this characteristic frequency the signal is suppressed in $|\tilde{\kappa}(f, \beta_\star, \lambda_\star)|^2$ relative to $|\tilde{\kappa}(f, \beta_{off}, \lambda_{off})|^2$. For a MBHB, the signal spreads across multiple frequencies, as frequency increases before the merger. Therefore, it is expected for the signal to be suppressed in a frequency band $(f_{min}, f_{max})$ where the frequencies $f_{min, max}$ depend on the parameters of the source. Indeed, Fig. \ref{fig:freq_mbhb} demonstrates that,  within a frequency band which roughly goes from $2 \cdot 10^{-4}$ to $10^{-2}$ Hz, the signal in $|\tilde{\kappa}(f, \beta_\star, \lambda_\star)|^2$ is suppressed by many order of magnitude relative to $|\tilde{\kappa}(f, \beta_{off}, \lambda_{off})|^2$.

\begin{figure}[ht!]
    \begin{subfigure}[b]{0.48\textwidth}
        \centering
        \includegraphics[scale=0.28]{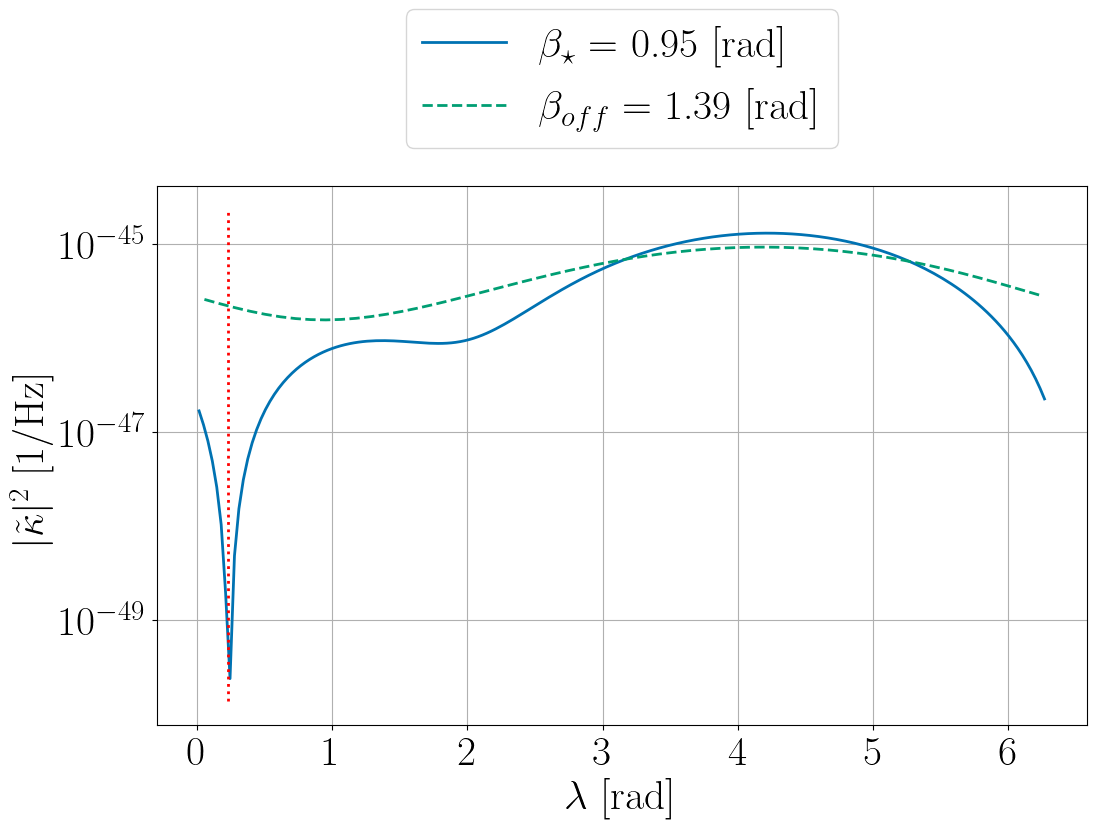}
        \caption{Coronagraphic TDI in frequency domain as a function of ecliptic latitude $\lambda$ at $f_0 = 3.788 $ mHz for VGB ZTFJ2243.}
        \label{fig:lambda_vgb}
    \end{subfigure}
    \hfill
    \begin{subfigure}[b]{0.48\textwidth}
        \centering
        \includegraphics[scale=0.28]{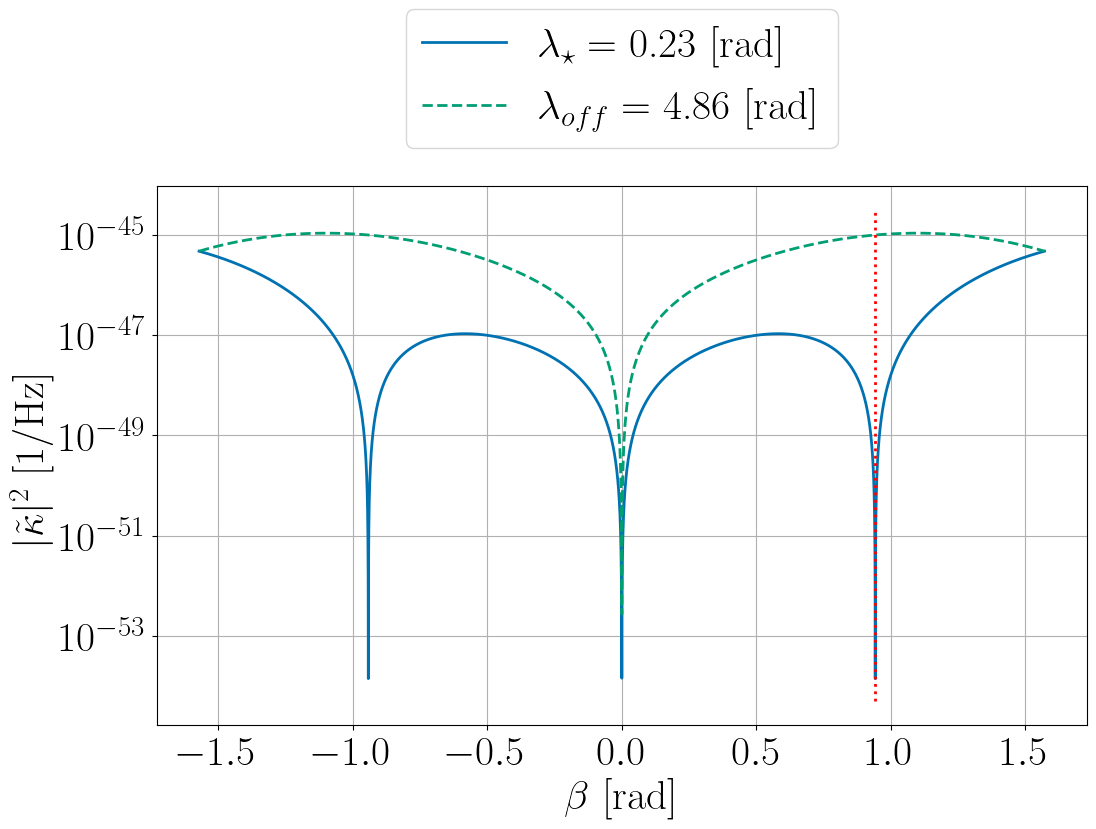}
        \caption{Coronagraphic TDI in frequency domain as a function of ecliptic longitude $\beta$ at $f_0 = 3.788 $ mHz for VGB ZTFJ2243.}
        \label{fig:beta_vgb}
    \end{subfigure}
    \hfill
    \begin{subfigure}[b]{0.48\textwidth}
        \centering
        \includegraphics[scale=0.28]{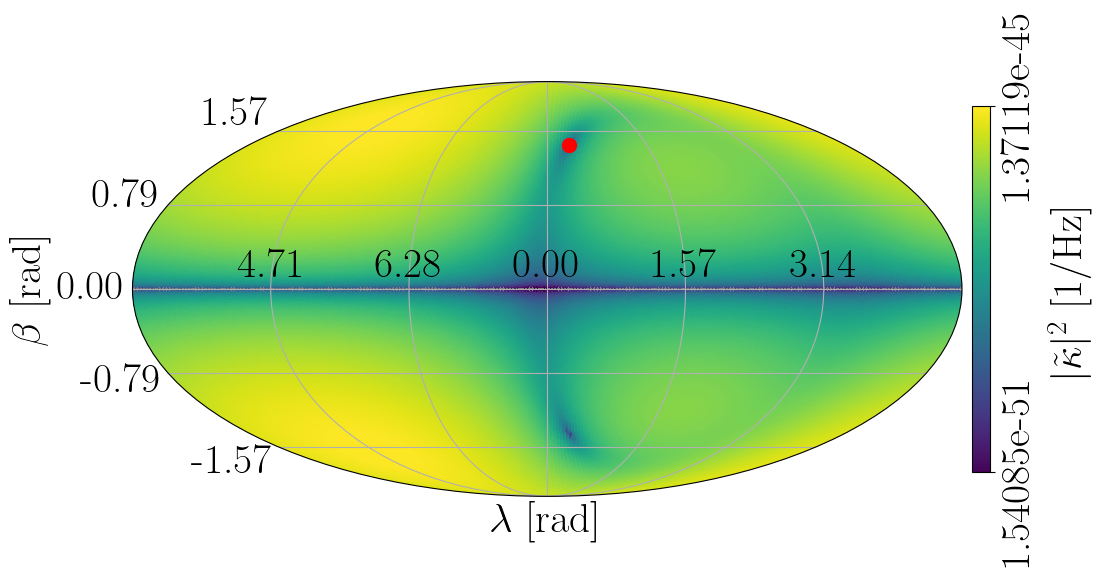}
        \caption{Coronagraphic TDI in frequency domain as a function of ecliptic latitude $\beta$ and ecliptic longitude $\lambda$ at $f_0 = 3.788 $ mHz for VGB ZTFJ2243. The red point indicates the known position of ZTFJ2243. The tiling of the sphere is done with \textsc{healpy} \footnote{See also \textsc{HEALPix}\cite{2005ApJ...622..759G}: http://healpix.sourceforge.net}\cite{Zonca2019} with $N_{side} = 64$.}
        \label{fig:map_vgb}
    \end{subfigure}
    \caption{Sky localization with coronagraphic TDI.}
    \label{fig:proof_sky}
\end{figure}

In Fig. \ref{fig:proof_sky} we show results for a scan of the sky at fixed frequency. First, similar to previous figures, in Fig. \ref{fig:lambda_vgb} and Fig. \ref{fig:beta_vgb} we compare two curves. In Fig. \ref{fig:lambda_vgb} we fix frequency and latitude and look at $\tilde{\kappa}$'s dependence on longitude. One curve is evaluated at $(f_0, \beta_\star)$ and another at $(f_0, \beta_{off})$. We see that $|\tilde{\kappa}(f_0, \beta_\star, \lambda)|^2$ is drastically reduced at $\lambda_\star$ while, although not flat, $|\tilde{\kappa}(f_0, \beta_{off}, \lambda)|^2$ does not show the same behavior. Second, an analogous profile is presented in Fig. \ref{fig:beta_vgb} where we look at the dependence on latitude instead. In this case, both curves vanish at $\beta=0$ rad, which corresponds to the plane of the constellation which is a blind spot of LISA. In fact, at this latitude the antenna pattern functions $\xi^\times_{ij}$ vanish and so does the vector $\vec{A}$ which enters the coronagraphic TDI calculation. Details are given in App. \ref{sec:app_latitude}. It is therefore to be expected for any signal to vanish at this latitude \cite{PhysRevD.70.062002}. Moreover, we notice that there is a degeneracy in $\beta$, which has been justified in Sec. \ref{sec:definition}. Finally, in Fig. \ref{fig:map_vgb} we present a scan of the sky in $|\tilde{\kappa}(f_0, \beta, \lambda)|^2$: the two darker spots in the sky correspond to a suppression of the signal by many orders of magnitude and are consistent with the known position of the source.

Lastly, it is important to stress that Fig. \ref{fig:map_vgb} is a function of frequency and that an similar sky map can be obtained as function of time. That being said, it is not as straight forward to analyze $\kappa$ in time domain. This is due to the fact that GW signals oscillate in time so at any $(\beta, \lambda)$ position there is a point in time where $\kappa$ is zero even though the signal is not suppressed. This can be observed in Fig. \ref{fig:time_vgb} and Fig. \ref{fig:time_mbhb} where the two curves intersect. In frequency domain however, provided a suitable frequency window $(f_{min}, f_{max})$ is selected, the values of $|\tilde{\kappa}|^2$ at different positions $(\beta, \lambda)$ can be compared more easily. For this reason, we stick to frequency domain. Nevertheless, it would be interesting to explore methods to analyze $\kappa$ in time domain in the future, possibly using time-frequency representations.

\section{Signal} \label{sec:signal}
 In this section we make the previous observations more quantitative by characterizing $\tilde{\kappa}$'s response to different types of signal. The characterization is done through a measure of angular resolution. We then evaluate how angular resolution is affected by different source parameters in the case of quasi-monochromatic sources such as GBs. Second, we test the developed technique on MBHBs, which are particularly relevant to low-latency searches.   

\subsection{Angular resolution} \label{sec:resolution}
In order to define the angular resolution of the method, it is necessary to quantify the suppression of the signal. It is important to note that at this stage, because there is no noise and no sampling involved, the procedure is completely deterministic. The adopted strategy here consists in defining a gain and then applying a threshold. To do so, a reference value $\tilde{\kappa}_0$ has to be chosen. This reference value is taken to be the value of $\tilde{\kappa}$ at the true position of the source. This value is then compared to the median value of all points $\tilde{\kappa}_i$ belonging to the circle of radius $\theta_i$ centered at $(\beta_\star, \lambda_\star)$. Thus, associated to the distance to the source $\theta_i$ is the gain
\begin{equation}
    g_i = 10 \cdot \log_{10} \frac{\mathrm{median}(|\tilde{\kappa_i}|^2)}{|\tilde{\kappa_0}|^2}.
\end{equation}
A threshold is then set to 3 dB, which means that at this distance the power has doubled. 

\begin{figure}[ht!]
    \centering
    \includegraphics[scale=0.28]{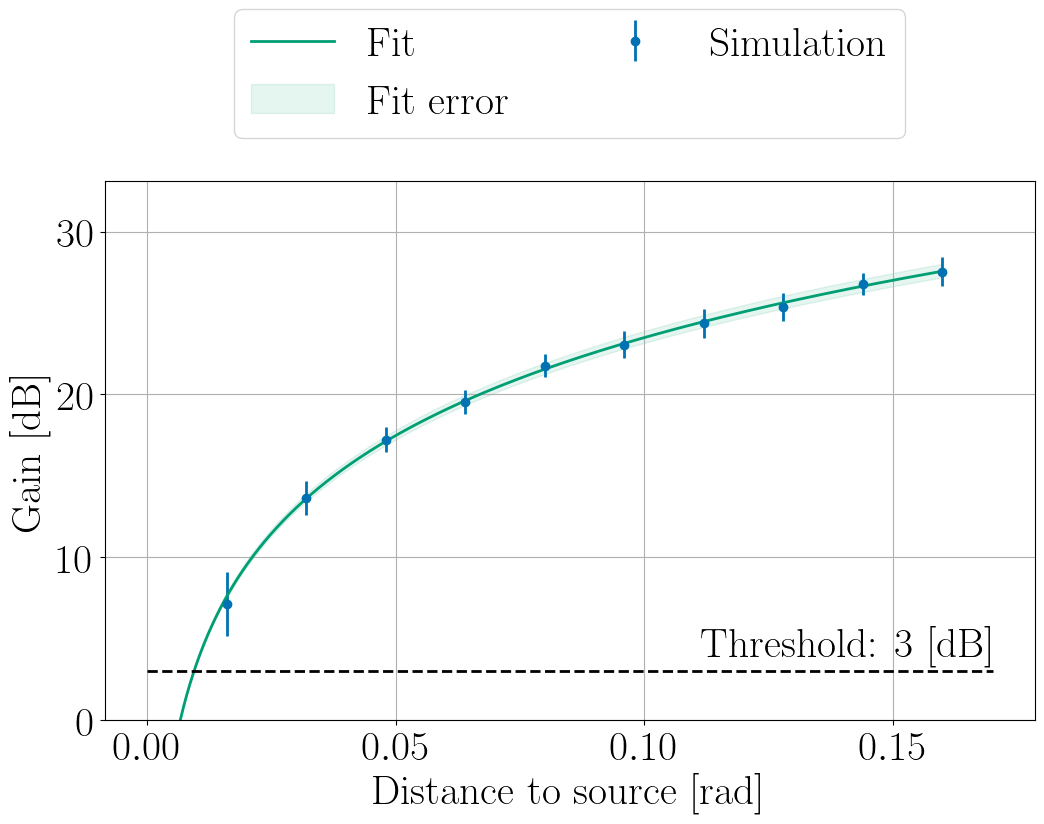}
    \caption{SDSSJ1908 fit by a logarithmic function $y=p_0 \log{x} + p_1$ where $p_0 = 8.68 \pm 0.11$ and $p_1 = 8.35 \pm 0.18$.}
    \label{fig:fit}
\end{figure}

This procedure is illustrated in Fig.\ref{fig:fit}. Starting from the scan of $|\tilde{\kappa}(f_0, \beta, \lambda)|^2$, showed in Fig.\ref{fig:map_vgb}, a series of gains $\{g_i\}$ is computed for different values of $\theta_i$ as described above. Error bars are computed from the median absolute deviation. The resulting points are fitted by an ad hoc logarithmic function so the points can be extrapolated to 3 dB. In this example, the distance at which the threshold is reached is $\theta_{3\mathrm{dB}} = (9.42 \pm 0.02) \cdot 10^{-3}$ rad. This angle can then be converted to a solid angle with 
\begin{equation}
    \Omega_{3\mathrm{dB}} = 2\pi(1-\cos{\theta_{3\mathrm{dB}}}) \times \frac{32400}{\pi^2} \, \mathrm{deg}^2.
\end{equation}

\subsection{Galactic binaries} \label{sec:gbs}
Having a measure of resolution, the proposed method for sky localization can be characterized in a quantitative manner. To this end, we apply the described technique to multiple datasets generated with the same software suite and under the same assumptions detailed in Sec.\ref{sec:demonstration}. The goal is to asses the capabilities of coronagraphic TDI in an idealized scenario. First, we consider the quasi-monochromatic case and investigate the influence of different parameters on resolution.

Commonly, GBs are parametrized by eight parameters, namely amplitude $\mathcal{A}$, frequency $f_0$, frequency derivative $\dot{f}$, inclination $\iota$, initial phase $\phi$, polarization $\Psi$, ecliptic latitude $\beta_\star$ and ecliptic longitude $\lambda_\star$. In what follows, we choose one VGB as a reference, here SDSSJ1908, and then create different datasets by modifying one parameter at a time. This way, the impact of each individual parameter can be evaluated. The parameters used in the simulations are summarized in Tab. \ref{tab:ref_vgb}. For each simulation, we compute $\tilde{\kappa}$ at the known frequency of the source $f_0$ and on a \textsc{HEALPix} grid $(\beta, \lambda)$ with $N_{side}=64$, which corresponds to 49152 pixels in total with pixel size $\theta_{\mathrm{pix}}=1.60\cdot 10^{-2}$ rad or $\Omega_{\mathrm{pix}}= 8.40 \cdot 10^{-1}\, \mathrm{deg}^2$ \cite{2005ApJ...622..759G}.

\begin{table}[ht!]
\caption{\label{tab:ref_vgb}%
Simulation parameters for GBs.
}
\begin{ruledtabular}
\begin{tabular}{ccccc}
\textrm{Parameter}&
\textrm{SDSSJ1908}&
\textrm{Range}&
\textrm{Units}&
\textrm{Points}\\
\colrule
$\mathcal{A}$ & $6.48 \cdot 10^{-23}$ & $[10^{-24}, 10^{-20}]$ & - & 100 \\
$f_0$ & 1.84 & $[1, 8]$ & mHz & 100 \\
$\dot{f}$ & $3.86 \cdot 10^{-18}$ & $[10^{-19}, 10^{-15}]$ & Hz/s & 100 \\
$\iota$ & $2.62 \cdot 10^{-18}$ & $[0,\pi]$ & rad & 100 \\
$\phi$ & 0 & $[0, 2\pi]$ & rad & 100 \\
$\Psi$ & 0 & $[0, 2\pi]$ & rad & 100 \\
$\beta_\star$ & $1.07$ & $(0, \frac{\pi}{2}]$ & rad & 1505 \footnote{Positions $(\beta_\star, \lambda_\star)$ are initially distributed on a \textsc{HEALPix} grid with $N_{side}=16$, which corresponds to $3072$ points. This number is then divided by half because we only consider the northern hemisphere. We further remove points at the equator and add a point at the pole.} \\
$\lambda_\star$ & $5.20$ & $[0, 2\pi]$ & rad & 1505 \footnotemark[1]
\end{tabular}
\end{ruledtabular}
\end{table}

\begin{figure}[ht!]
    \begin{subfigure}[b]{0.48\textwidth}
        \centering
        \includegraphics[scale=0.28]{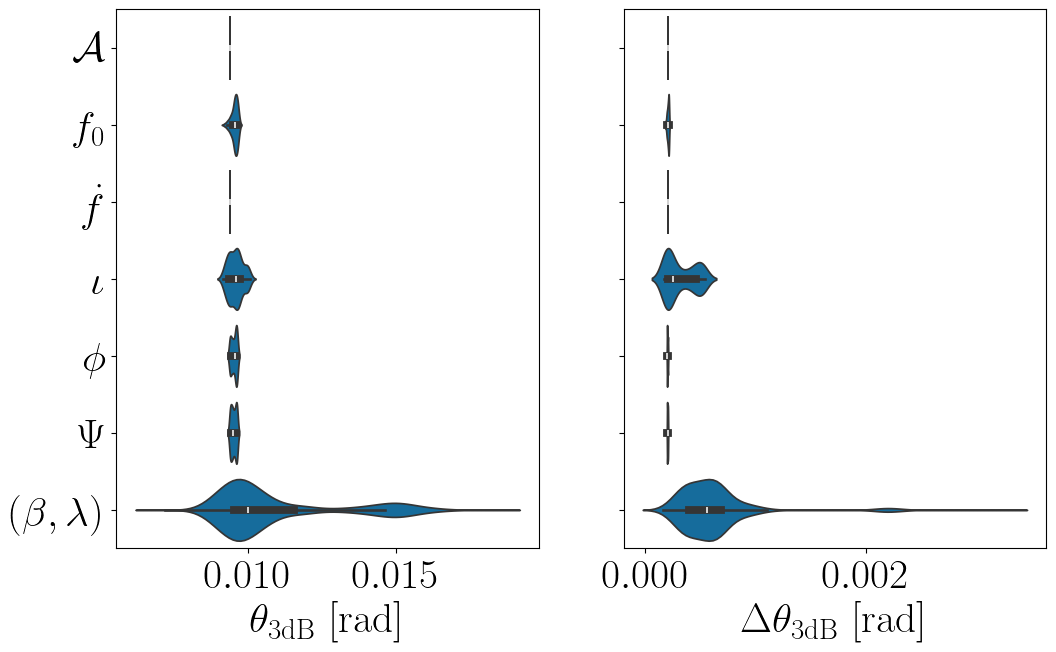}
        \caption{For each dataset described in Tab. \ref{tab:ref_vgb}, distance to source at 3 dB (left) and associated error (right) distributions.}
        \label{fig:violin_vgb}
    \end{subfigure}
    \hfill
    \begin{subfigure}[b]{0.48\textwidth}
        \centering
        \includegraphics[scale=0.28]{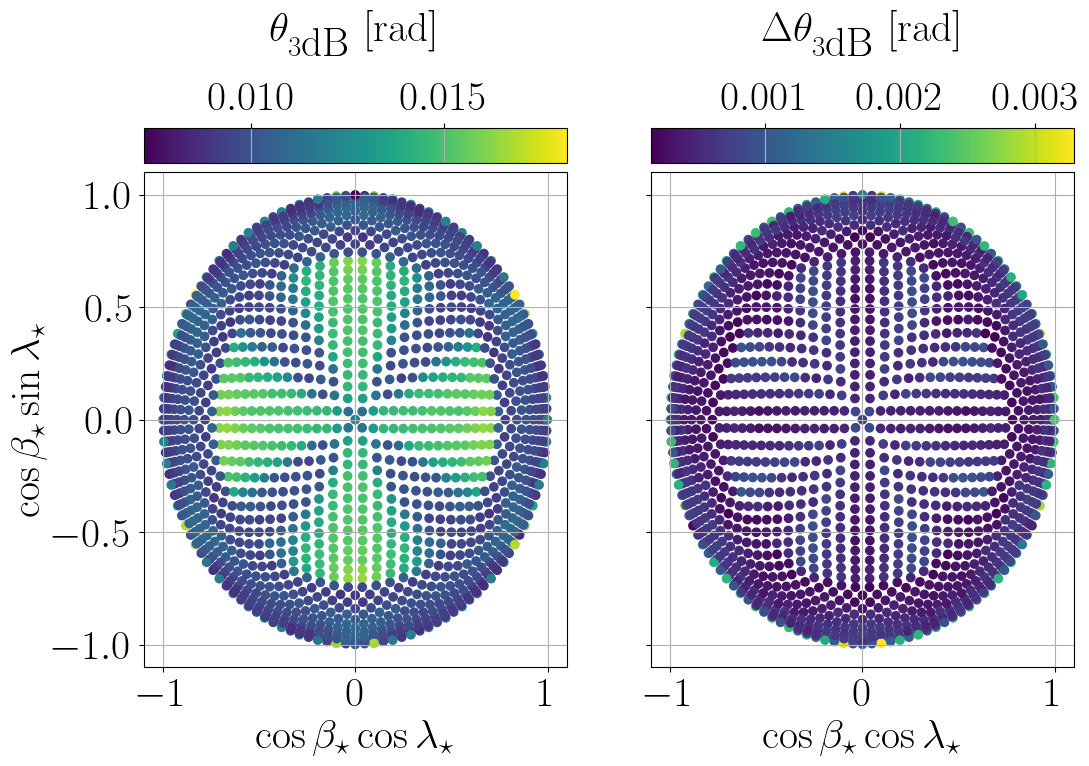}
        \caption{Northern hemisphere projection. Each point corresponds to a simulated signal. Its position is the position of the source in the simulation and the color map represents distance to source at 3 dB (left) and associated error (right).}
        \label{fig:position_vgb}
    \end{subfigure}
    \caption{Results for GBs simulations.}
    \label{fig:results_vgb}
\end{figure}

Results are summarized in Fig. \ref{fig:violin_vgb} where the distributions of $\theta_{3\mathrm{dB}}$ and associated error are presented for each dataset. Mean values and standard deviations are reported in App. \ref{sec:app_signal} in $\mathrm{deg}^2$. Although not the main focus of this section, mean values can be put into perspective by comparing them to the resolutions obtained for other VGBs. This information is also provided in App. \ref{sec:app_signal}.

In light of Fig. \ref{fig:violin_vgb}, we can split the set of eight parameters in three groups: those that have no impact, those that have a moderate impact and those that have a high impact on the proposed sky localization method. In the first group are parameters $\mathcal{A}$ and $\dot{f}$. In fact, given that the simulations are noiseless, it is to be expected that the amplitude has no impact on the method. Moreover, because simulations last for less than a week, it is also expected for drift in frequency not to affect resolution. In the second group are parameters $\{f_0, \iota, \phi, \Psi\}$ for which the observed distributions can be given a physical explanation if having in mind the waveform formula for GBs. Lastly, in the third group is the position of the source. Not only is the distribution more spread, it also has two peaks. Mapping the position of the source to its resolution, we can see in Fig. \ref{fig:position_vgb} that the points with higher $\theta_{3\mathrm{dB}}$ form a cross-like pattern close to the pole. There is no physical interpretation to this pattern, since a cross-like pattern is not consistent with the symmetries of LISA, which is a triangle. It is likely an artifact from the method used to discretize the sphere. 

Regardless, the parameter which has the most significant impact on resolution is the position of the source. In fact, $\theta_{3\mathrm{dB}}$ varied from $7.21 \cdot 10^{-3}$ to $1.82 \cdot 10^{-2}$ rad, that is from $5.36 \cdot 10^{-1}$ to $3.41 \, \mathrm{deg}^2$ depending on the position of the source. Moreover, one should keep in mind that this applies to the position of the source with respect to the constellation. For instance, with realistic orbits a source close to the ecliptic plane will not necessarily be close to the plane of the constellation, which is the blind spot of this method. 

\subsection{Massive black hole mergers} \label{sec:mbhbs}
The MBHB case shares similarities with the GB case, so instead of going through all MBHB parameters and analyzing their impact on resolution we choose to analyze all of 15 MBHBs from the Sangria dataset. The main challenge here is to deal with the polychromatic nature of the signal.

The proposed method consists in analyzing each frequency $f_i$ within a frequency band $(f_{min}, f_{max})$ with the technique introduced in Sec. \ref{sec:resolution}. It goes as follows:
\begin{enumerate}
    \item Compute $\tilde{\kappa}(f_i, \beta, \lambda)$ for $(\beta, \lambda)$ in a \textsc{HEALPix} grid of $N_{side}=64$. This step corresponds to producing a sky map similar to Fig. \ref{fig:map_vgb}.
    \item Estimate the angular resolution $\theta_{3\mathrm{dB}}$ for this sky map with the procedure detailed in Sec. \ref{sec:resolution}. At this step we compute the gain and fit it by a function $y = p_0 \log{x} + p_1$. The results at this step are resolution $\theta_{3\mathrm{dB}}$ and fit parameters $p_0$ and $p_1$.
    \item Exclude unphysical results. We use two criteria to determine if a result is physical:
    \begin{itemize}
        \item $\theta_{3\mathrm{dB}} \in (0, 0.175)$ rad: if the estimated resolution $\theta_{3\mathrm{dB}}$ is outside of this range we consider that the algorithm was unsuccessful.
        \item $p_0 > 0$: signal suppression translates into a decrease in gain when approaching the source (see Fig \ref{fig:fit}). However, with the chosen fit function if $p_0 \leq 0$ then the gain increases when approaching the source. Thus, we consider that the chosen function only fits the data when $p_0 > 0$.
    \end{itemize}
\end{enumerate}

Analogous to GBs, simulations are generated with the software and under the assumptions presented in Sec. \ref{sec:demonstration}. Details about MBHBs parameters can be found in the LDC documentation \cite{sangria}, but a selection of parameters are given in App. \ref{sec:app_signal}. Finally, we consider 1000 frequencies $f_i$ distributed logarithmically between $4 \cdot 10^{-5}$ and $2 \cdot 10^{-2}$ Hz for all sources.

\begin{figure}[ht!]
    \centering
    \includegraphics[scale=0.28]{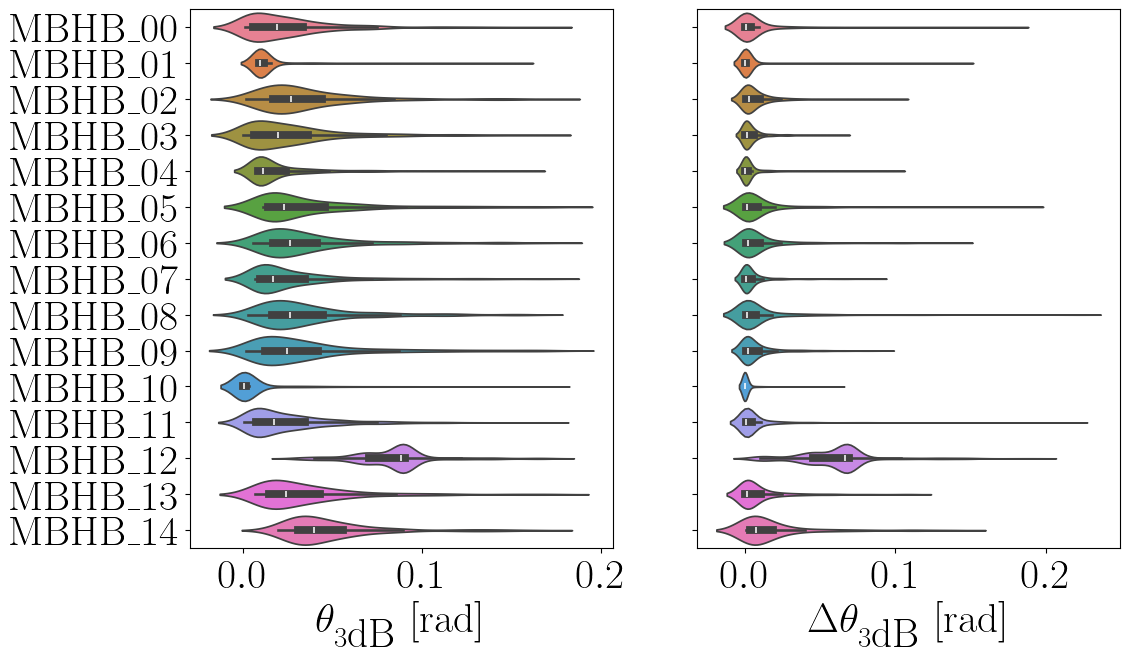}
    \caption{Results for MBHBs simulations. For each MBHB in the Sangria dataset, distance to source at 3 dB (left) and associated error (right) distributions.}
    \label{fig:violin_mbhb}
\end{figure}

Results are summarized in Fig. \ref{fig:violin_mbhb} where the $\theta_{3\mathrm{dB}}$ and associated error distributions are plotted for every MBHB in the Sangria dataset. We can note that all distributions have a similar shape, being spread and having a long tail. This response deserves to be carefully studied in the future, in particular when designing a source localization algorithm based on $\tilde{\kappa}$. Nevertheless, on average, distributions have mean value $(3.38 \pm 1.14) \cdot 10^{-2}$ rad, which corresponds to $(11.4 \pm 7.8) \, \mathrm{deg}^2$. The best result is achieved for MBHB\_10 with mean value $(8.17 \pm 1.7) \cdot 10^{-3}$ rad, that is $(6.89 \pm 3.01) \, \mathrm{deg}^2$. For completeness, mean values and standard deviations for each distribution are provided in App. \ref{sec:app_signal} in $\mathrm{deg}^2$.

The goal of this study is to demonstrate the feasibility of such an analysis on polychromatic sources. This result shows that coronagraphic TDI can be used to determine the position of polychromatic sources such as MBHBs.

\section{Detector response} \label{sec:detector}
Coronagraphic TDI and detector response are intertwined by construction. It is therefore fundamental to the characterization of the method to investigate $\tilde{\kappa}$'s response in different scenarios not only in terms of signal but also in terms of instrument. First, we analyze $\tilde{\kappa}$'s response to a simple analytic noise model. Then, because the computation of $\vec{A}$ relies on the knowledge of arm lengths and spacecraft position, we study $\tilde{\kappa}$'s response when making different assumptions on the position of the LISA constellation.

\subsection{Instrument noise} \label{sec:noise}
In frequency domain, the response of $\tilde{\kappa}$ to noise can be estimated with an analytic noise model. For a given TDI variable, an analytic noise model consists in a TDI transfer function and a physically-motivated spectral shape. In a simplified noise model, two sources of secondary noise are taken into account: test mass (TM) acceleration and optical metrology system (OMS) noise. Assuming noise amplitudes are identical in all spacecraft, TM and OMS spectral shapes, $S^{\mathrm{TM}}$ and $S^{\mathrm{OMS}}$ respectively, are given by the following expressions \cite{LISA}:
\begin{widetext}
\begin{align}
    S^{\mathrm{TM}}(f) &= A^2 \times \Big[1+\Big(\frac{0.4 \times 10^{-3} \mathrm{Hz}}{f}\Big)^2\Big] \times \Big[1+\Big(\frac{f}{8 \times 10^{-3} \mathrm{Hz}}\Big)^4\Big] \times \Big(\frac{1}{\omega c}\Big)^2 \times (\mathrm{m}^2/\mathrm{s}^3) \\
    S^{\mathrm{OMS}}(f) &= P^2 \times \Big[1+\Big(\frac{2 \times 10^{-3} \mathrm{Hz}}{f}\Big)^2\Big] \times \Big(\frac{\omega}{c}\Big)^2 \times (\mathrm{m}^2/\mathrm{Hz}),
    \label{eq:spectral_shape}
\end{align}      
\end{widetext}  
where $A = 3 \cdot 10^{-15} \, \mathrm{m}/(\mathrm{s}^2\sqrt{\mathrm{Hz}})$ , $P = 15 \cdot 10^{-12} \, \mathrm{m}/\sqrt{\mathrm{Hz}}$ and $\omega = 2 \pi f$. For simplicity, we assume equal arms so when applying the associated transfer functions one gets the power spectral densities (PSDs) \cite{PhysRevD.107.123531} 
\begin{align}
    S_{\alpha\alpha} &= S_{\beta\beta} = S_{\gamma\gamma} \nonumber \\
    &= 6S^{\mathrm{OMS}} \nonumber \\
    &\quad + 4 \times [3 - 2\cos(\omega L) - \cos(3\omega L)] \times S^{\mathrm{TM}}
\end{align}
and cross spectral densities (CSDs) \cite{PhysRevD.107.123531}
\begin{align}
    S_{\alpha\beta} &= S_{\alpha\gamma} = S_{\beta\gamma} \nonumber \\
    &= 2 \times [2\cos(\omega L) + \cos(2 \omega L)] \times S^{\mathrm{OMS}} \nonumber \\
    &\quad- 4 \times [1 - 2\cos(\omega L) - \cos(3\omega L)] \times S^{\mathrm{TM}}
\end{align}
for Sagnac variables $(\tilde{\alpha}, \tilde{\beta}, \tilde{\gamma})$. These are the building blocks for defining an analytic noise model for coronagraphic TDI. In fact, because $\tilde{\kappa}$ is a combination of Sagnac TDI variables, it suffices to use the same transformation, given in Eq.\eqref{eq:def}, to derive an analytic noise model for $\tilde{\kappa}$:
\begin{align}
    S_{\kappa\kappa} & = \langle \tilde{\kappa} \tilde{\kappa}^* \rangle \nonumber \\
    & = \begin{pmatrix}
        a_\alpha & a_\beta & a_\gamma
    \end{pmatrix} \begin{pmatrix}
        S_{\alpha\alpha} & S_{\alpha\beta} & S_{\alpha\gamma} \\
        S_{\beta\alpha} & S_{\beta\beta} & S_{\beta\gamma} \\
        S_{\gamma\alpha} & S_{\gamma\beta} & S_{\gamma\gamma} 
    \end{pmatrix} \begin{pmatrix}
        a_\alpha^* \\
        a_\beta^* \\
        a_\gamma^*
    \end{pmatrix}. \label{eq:noise}
\end{align}

\begin{figure}[ht!]
    \centering
    \includegraphics[scale=0.28]{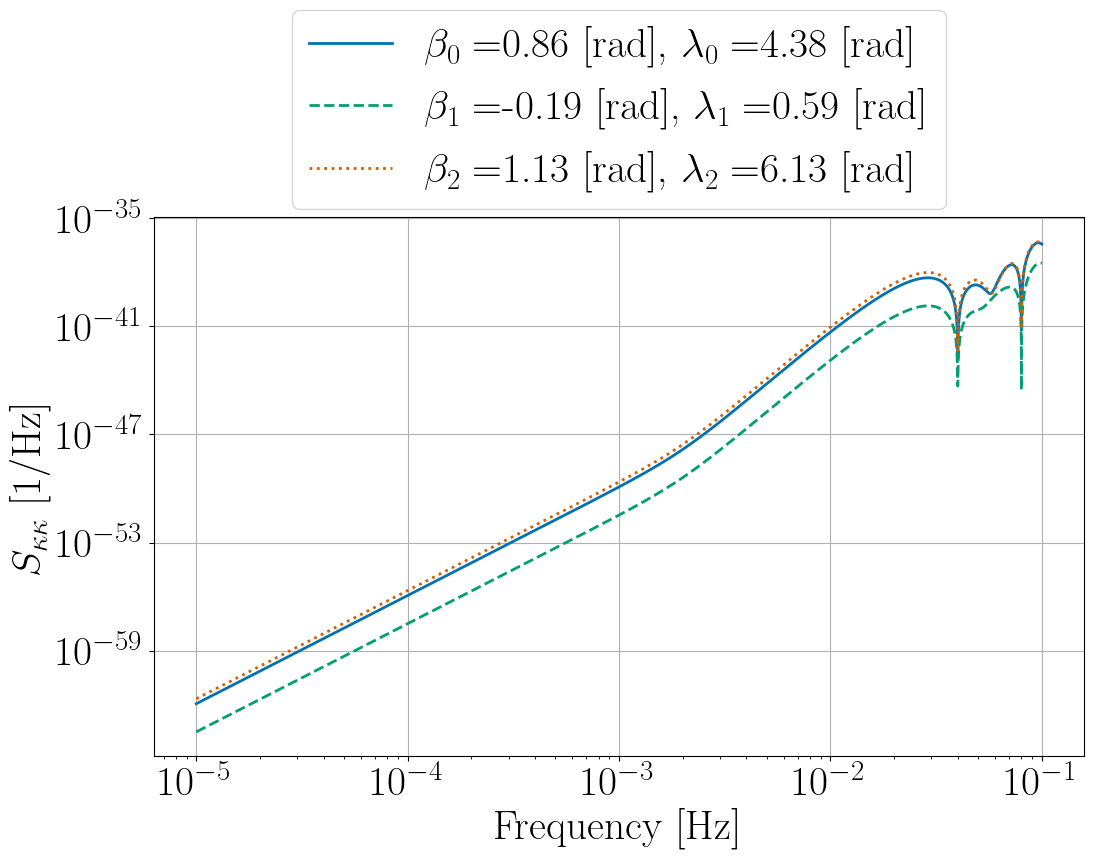}
    \caption{Coronagraphic noise spectrum for different positions based on model Eq.\eqref{eq:noise}.}
    \label{fig:noise_freq}
\end{figure}

It is important to notice that the analytic noise model for $\tilde{\kappa}$ depends on frequency and sky position $(\beta, \lambda)$. The dependence of $S_{\kappa\kappa}$ on frequency is presented in Fig. \ref{fig:noise_freq}, where $S_{\kappa\kappa}(f, \beta, \lambda)$ is plotted for three fixed positions $(\beta_k, \lambda_k)$ with $k \in \{0, 1, 2\}$. This figure is analogous to Fig. \ref{fig:proof_freq} and states what the overall shape of $S_{\kappa\kappa}$ is. Notice that the shape of the noise spectrum is roughly the same regardless of the position  $(\beta, \lambda)$. The noise level however, is lower for $(\beta_1, \lambda_1)$ because this position is closer to the plane of constellation, a region where $\tilde{\kappa}$ tends to zero because vector $\vec{A}$ vanishes, as discussed in Sec. \ref{sec:tdi} and detailed in App. \ref{sec:app_latitude}. In fact, at latitude zero $S_{\kappa\kappa}(f, 0, \lambda)$ is exactly zero for any $f$ and any $\lambda$.

\begin{figure}[ht!]
    \begin{subfigure}[b]{0.48\textwidth}
        \centering
        \includegraphics[scale=0.28]{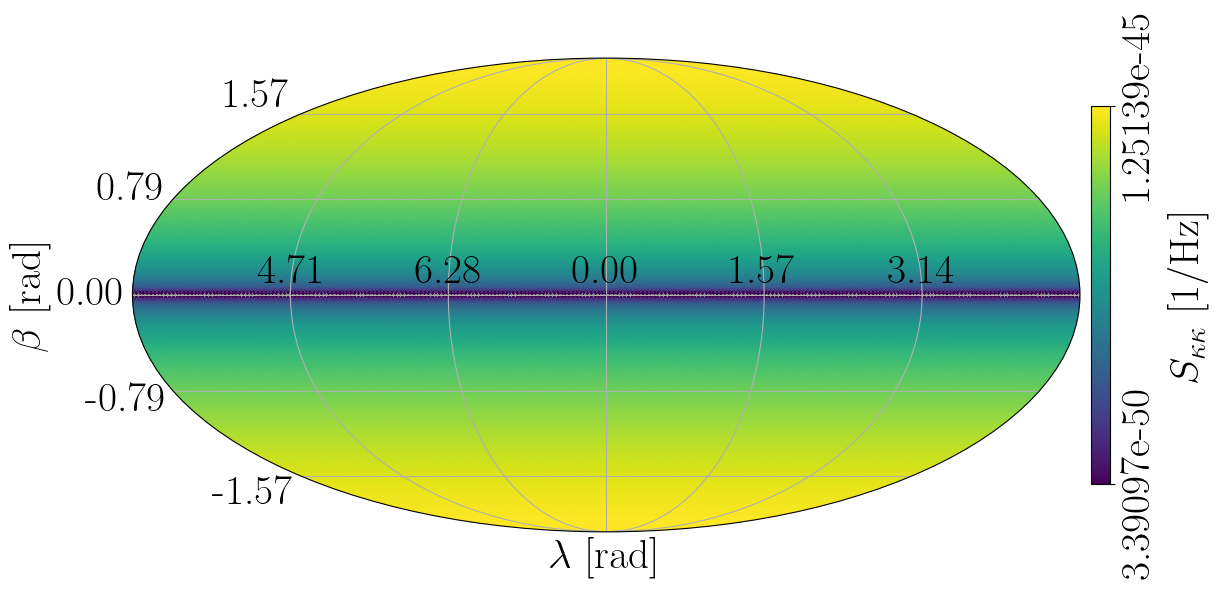}
        \caption{$S_{\kappa\kappa}$ at $f_0 = 3.788 $ mHz. The tiling of the sphere is done with $N_{side}=64$.}
        \label{fig:noise_map}
    \end{subfigure}
    \hfill
    \begin{subfigure}[b]{0.48\textwidth}
        \centering
        \includegraphics[scale=0.28]{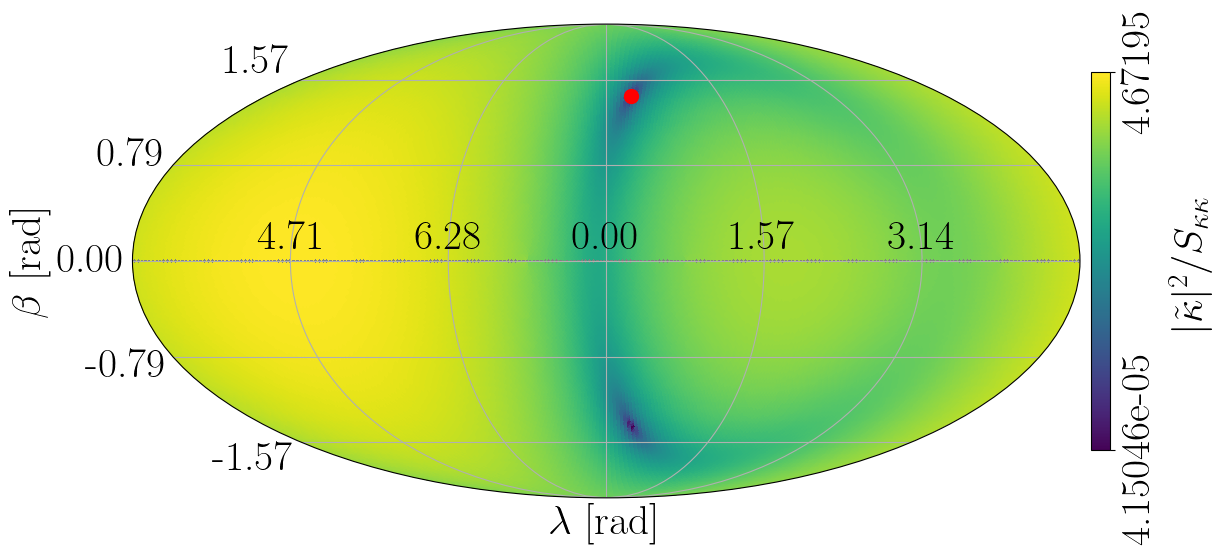}
        \caption{Coronagraphic TDI over $S_{\kappa\kappa}$ in frequency domain as a function of ecliptic latitude $\beta$ and ecliptic longitude $\lambda$ at $f_0 = 3.788 $ mHz for VGB ZTFJ2243. The tiling of the sphere is done with $N_{side}=64$.}
        \label{fig:noise_snr}
    \end{subfigure}
    \hfill
    \caption{Noise and sky localization with coronagraphic TDI.}
    \label{fig:noise_sky}
\end{figure}

Furthermore, Fig. \ref{fig:noise_map} exhibits a sky map of $S_{\kappa\kappa}(f, \beta, \lambda)$ at fixed frequency $f_0$. It is clear from Fig. \ref{fig:noise_map} that the value of $S_{\kappa\kappa}$ decreases when approaching the plane of the constellation at $\beta=0$ rad. What is more, Fig. \ref{fig:noise_map} shows that $S_{\kappa\kappa}(f, \beta, \lambda)$ does not depend on longitude $\lambda$. This is not evident from the mathematical derivation of $\tilde{\kappa}$ but makes physical sense: $S_{\kappa\kappa}$ is essentially the response of the detector seen through the lenses of coronagraphic TDI and, by construction, the plane of the constellation is the only blind region of $\tilde{\kappa}$. No particular dependence on $\lambda$ is to be expected.

Finally, we compute the quantity $|\tilde{\kappa}|^2/S_{\kappa\kappa}$ on the entire sphere at fixed frequency $f_0$. The resulting sky map is presented in Fig. \ref{fig:noise_snr}. It is the ratio between Fig. \ref{fig:map_vgb} and Fig. \ref{fig:noise_map}. Remarkably, the apparent signal suppression around $\beta = 0$ rad in Fig. \ref{fig:map_vgb} is no longer visible in Fig. \ref{fig:noise_snr}. Indeed, by dividing by $S_{\kappa\kappa}$ we correct for an effect due to the detector's response in the coronagraphic TDI channel. This result opens the way for future analyses which further investigate the behavior of $\tilde{\kappa}$ beyond the noiseless case studied in this paper.

\subsection{LISA constellation} \label{sec:constellation}
Regarding the position of the LISA constellation we have made two assumptions thus far. First, we have assumed orbits to be static. As already mentioned, this is not true in reality but is a valid approximation for relatively short periods of time. Second, we have assumed the constellation to lie in the ecliptic plane. Yet, LISA will follow a cartwheel motion around the Sun, trailing the Earth \cite{LISA}. For this reason, we challenge this second assumption by shifting towards realistic LISA orbits provided by the European Space Agency (ESA) \cite{EsaOrbits}. 

In this dataset, we generate 52 simulations of a given VGB, here ESCet. Over one year, assuming LISA evolves according to ESA orbits, we take a picture of the LISA constellation every week and then simulate a VGB signal with static orbits. All simulations are noiseless and last for 5.8 days.

From a coronagraphic TDI perspective, this shift towards realistic orbits has two consequences. On the one hand, because coronagraphic TDI is constructed from the detector’s response, it necessarily points at a GW source in the reference frame of LISA. That is, the position $(\beta, \lambda)$ where signal is suppressed in the coronagraphic TDI channel is given in the LISA frame. However, this information has to be communicated to the broader astronomical community in solar system barycentric (SSB) ecliptic coordinates. Therefore, one has to translate results from the LISA frame to the SSB frame. This is achieved by defining a rotation matrix which links the two frames. For conciseness, the derivation of this rotation matrix is left to App. \ref{sec:app_rotation}. A similar derivation can also be found in \cite{PhysRevD.68.122001}. So far both frames overlapped so this step was not required. Now that LISA evolves in time so does the apparent position of the source in the LISA frame.

On the other hand, the position of the LISA constellation directly enters the computation of $\tilde{\kappa}$. Indeed, $\vec{A}$ is computed for some spacecraft position $\vec{x}_i$ with $i \in \{1, 2, 3\}$ and for some arm length $L_{ij}$. Likewise, the position of the LISA constellation also affects the computation of the standard TDI variables carried by vector $\vec{D}$. So, for clarity, we introduce the notations
\begin{equation}
    \vec{A}_{\vec{x}_i(t_n), L_{ij}(t_n)} \quad \text{and} \quad \vec{D}_{\vec{x}_i(t_n), L_{ij}(t_n)}
\end{equation}
where $t_n$ is the time where the snapshot of the constellation is taken and $n$ ranges from 0 to 51. In what follows we refer to $t_n$ as snapshot time, not to be confused with integration time which is kept constant at 5.8 days. Then, for a data vector $\vec{D}_{\vec{x}_i(t_n), L_{ij}(t_n)}$, we distinguish two cases:
\begin{enumerate}
    \item Synchronous case: $\tilde{\kappa}$ is computed with an updated vector $\vec{A}_{\vec{x}_i(t_n), L_{ij}(t_n)}$. In this case we are interested in the evolution of $\tilde{\kappa}$'s sky localization performance as LISA moves around the Sun.
    \item Asynchronous case: $\tilde{\kappa}$ is computed with a constant vector $\vec{A}_{\vec{x}_i(t_0), L_{ij}(t_0)}$. In this case we are interested in determining the duration of validity of coefficients in $\vec{A}$.
\end{enumerate}
Results are presented and discussed in the next section.

\subsection{Source tracking} \label{sec:tracking}
Because LISA now evolves in time, we need to track the position of the source in the LISA frame as seen through the lenses of coronagraphic TDI. So for every time $t_n$ we identify the point where $\tilde{\kappa}$ is minimized. This point corresponds to the estimated position of the source. This is done by first detouring a region of interest around an initial guess and then finding a local minimum in this region. This is done in the two cases described above, namely synchronous and asynchronous case. Then angular resolution $\theta_{3\mathrm{dB}}$ is computed as described in Sec. \ref{sec:resolution}, but taking as reference value $\tilde{\kappa}_0$, the value of $\tilde{\kappa}$ at the identified local minimum.

\begin{figure}[ht!]
    \begin{subfigure}[b]{0.48\textwidth}
        \centering
        \includegraphics[scale=0.28]{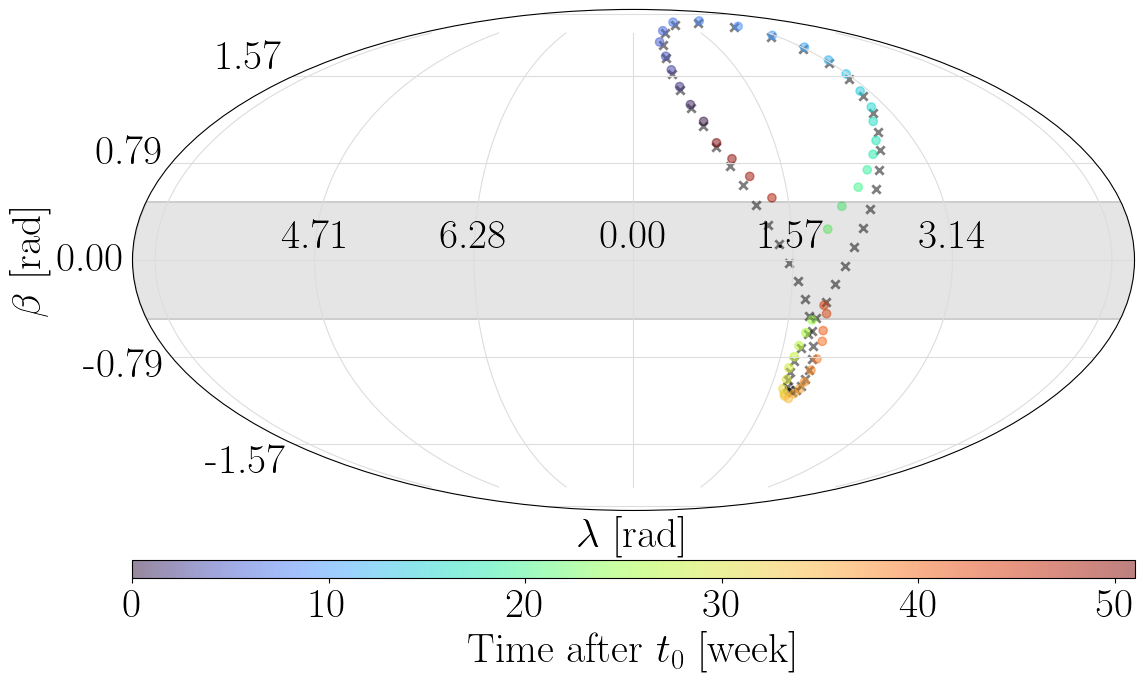}
        \caption{Synchronous case.}
        \label{fig:sync_map}
    \end{subfigure}
    \hfill
    \begin{subfigure}[b]{0.48\textwidth}
        \centering
        \includegraphics[scale=0.28]{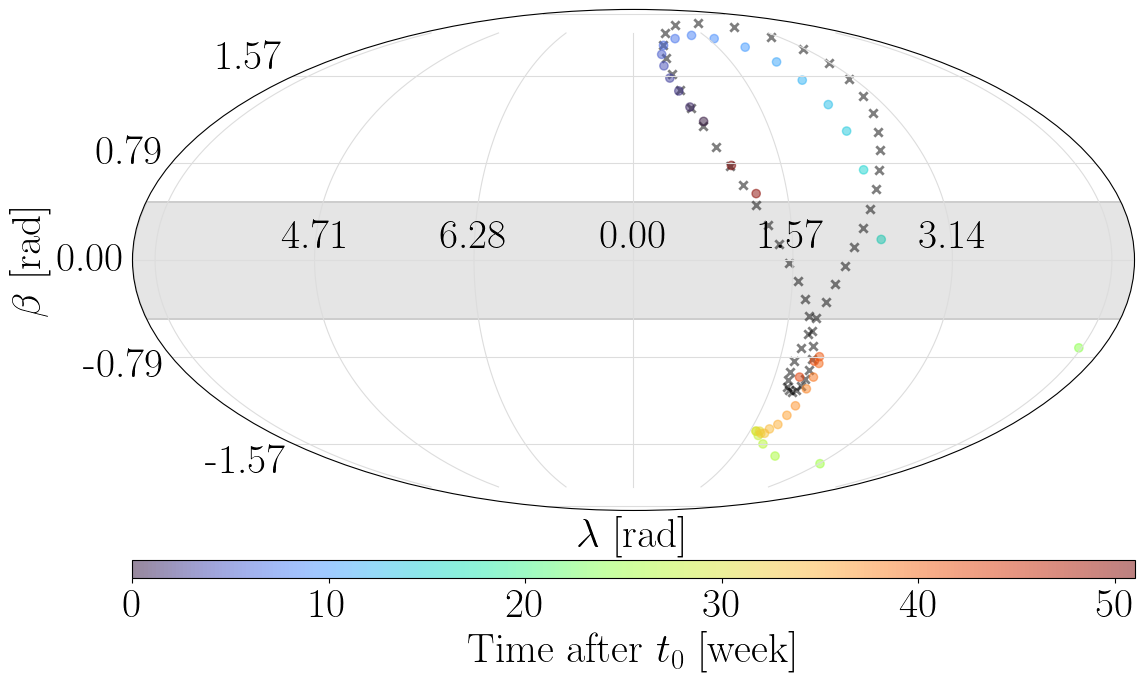}
        \caption{Asynchronous case.}
        \label{fig:async_map}
    \end{subfigure}
    \caption{Source tracking. Dots represent the source position obtained from the analysis of $\tilde{\kappa}$, while crosses represent the known position of ESCet in the LISA frame at time $t_n$. The gray area corresponds to $|\beta| < 0.3$ rad.}
    \label{fig:tracking}
\end{figure}

In the synchronous case, Fig. \ref{fig:sync_map}, $\tilde{\kappa}$ successfully tracks the position of the source in the LISA frame as long as the latter is away from the plane of the constellation. In the initial case treated in  Sec. \ref{sec:gbs}, the area around which $\tilde{\kappa}$ was not able to detect a GW source was much narrower. The fact that this area is broader in the present case is an effect due to the motion of LISA and in particular of the plane of the constellation. In the asynchronous case, however, $\tilde{\kappa}$ is not able to correctly track the position of the source in the LISA frame after a couple months, as seen in Fig. \ref{fig:async_map}. This is to be expected since after this period of time the LISA constellation configuration has considerably changed.

\begin{figure}[ht!]
    \begin{subfigure}[b]{0.48\textwidth}
        \centering
        \includegraphics[scale=0.28]{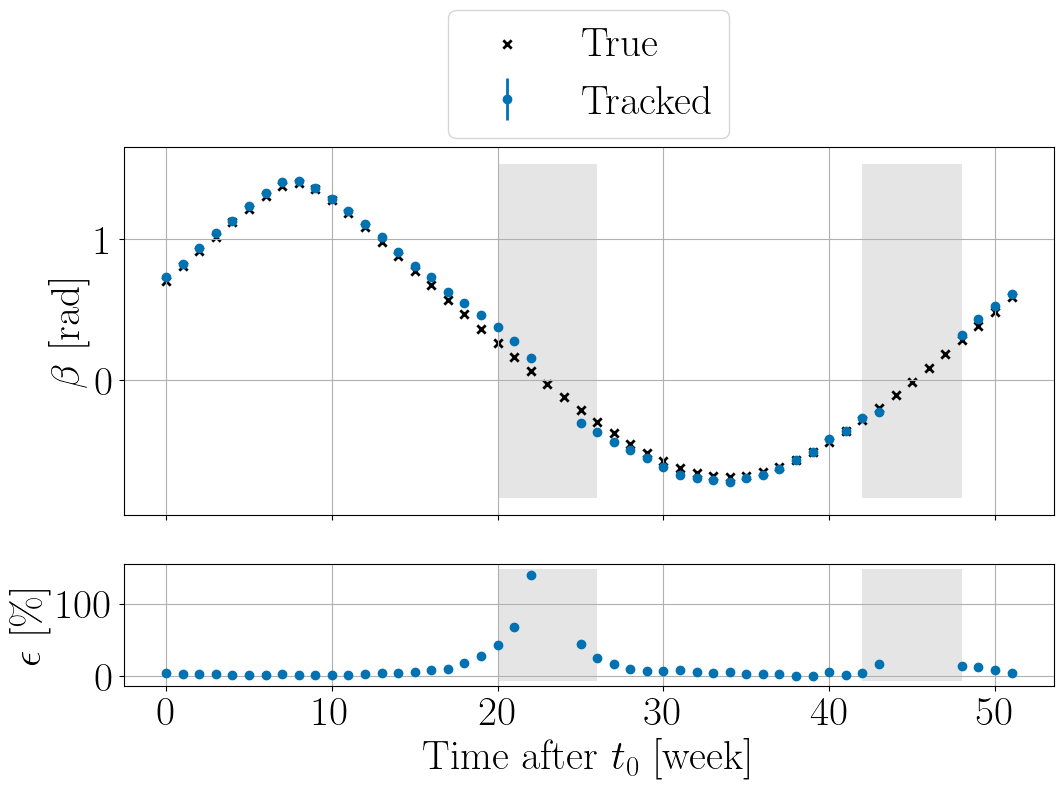}
        \caption{Source latitude $\beta$ obtained from the analysis of $\tilde{\kappa}$ (tracked), known latitude (true) and absolute relative error ($\epsilon$) as a function of snapshot time.}
        \label{fig:sync_beta}
    \end{subfigure}
    \hfill
    \begin{subfigure}[b]{0.48\textwidth}
        \centering
        \includegraphics[scale=0.28]{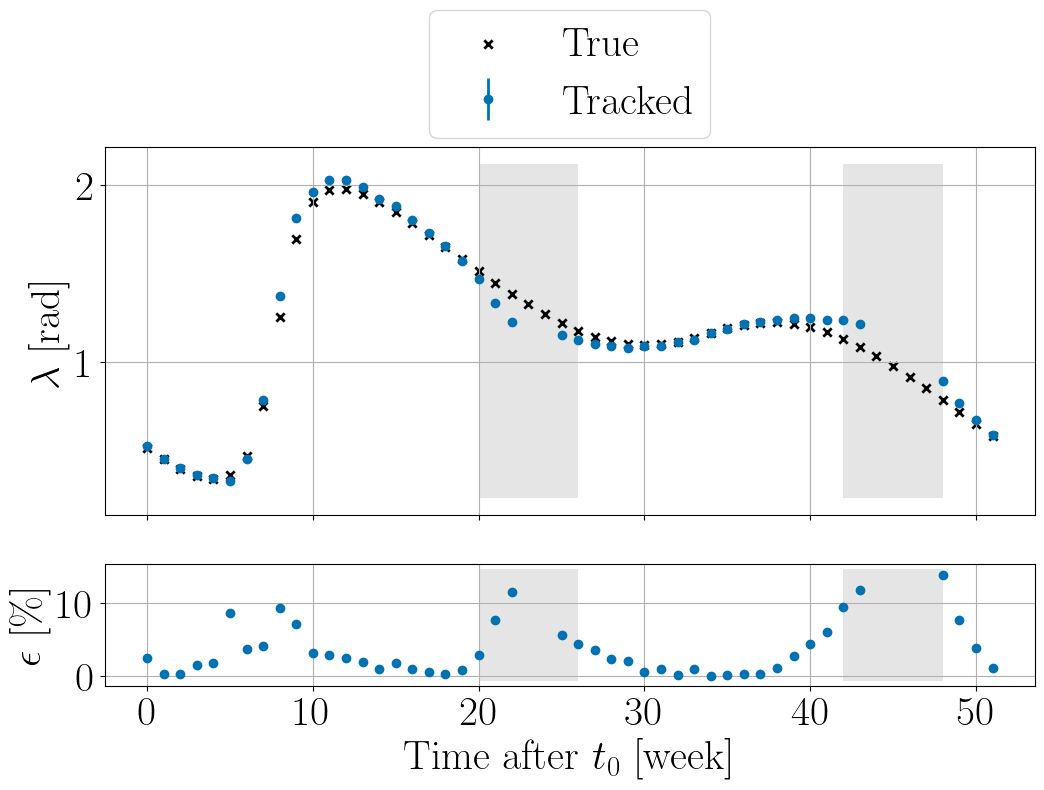}
        \caption{Source longitude $\lambda$ obtained from the analysis of $\tilde{\kappa}$ (tracked), known longitude (true) and absolute relative error ($\epsilon$) as a function of snapshot time.}
        \label{fig:sync_lambda}
    \end{subfigure}
    \caption{Synchronous case: latitude and longitude estimation as a function of snapshot time. The gray area corresponds to the time when the apparent latitude of the source in the LISA frame is of $|\beta| < 0.3$ rad.}
    \label{fig:sync_time}
\end{figure}

\begin{figure}[ht!]
    \begin{subfigure}[b]{0.48\textwidth}
        \centering
        \includegraphics[scale=0.28]{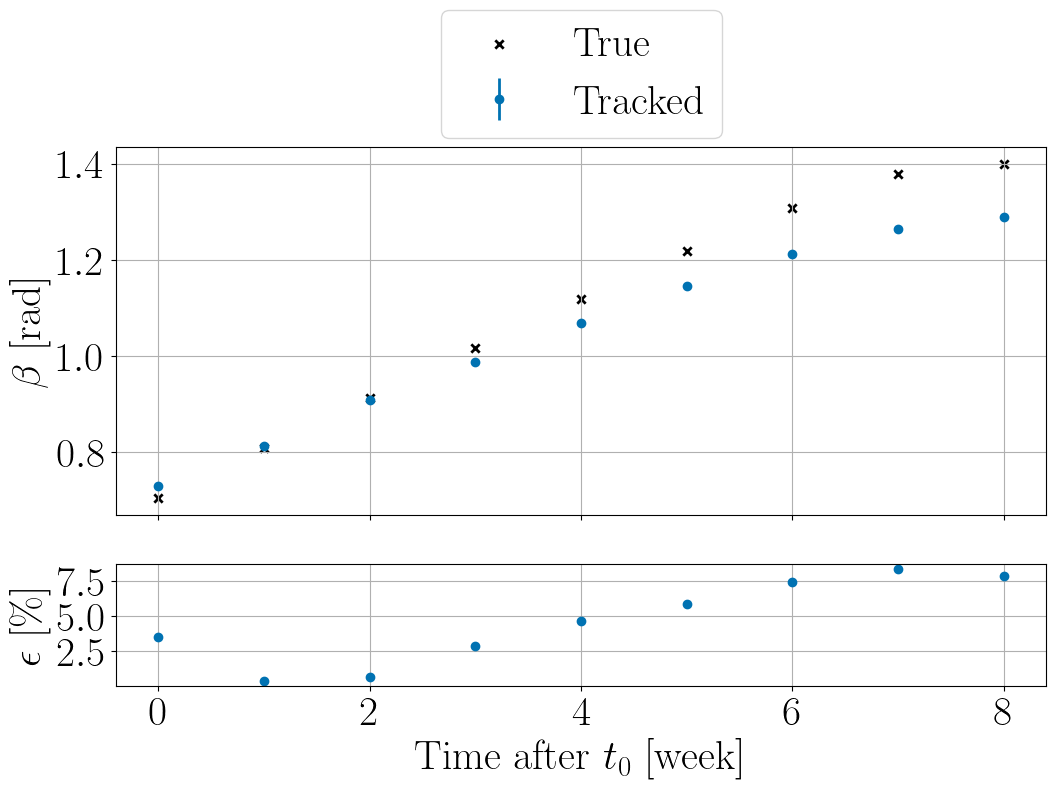}
        \caption{Source latitude $\beta$ obtained from the analysis of $\tilde{\kappa}$ (tracked), known latitude (true) and absolute relative error ($\epsilon$) as a function of snapshot time.}
        \label{fig:async_beta}
    \end{subfigure}
    \hfill
    \begin{subfigure}[b]{0.48\textwidth}
        \centering
        \includegraphics[scale=0.28]{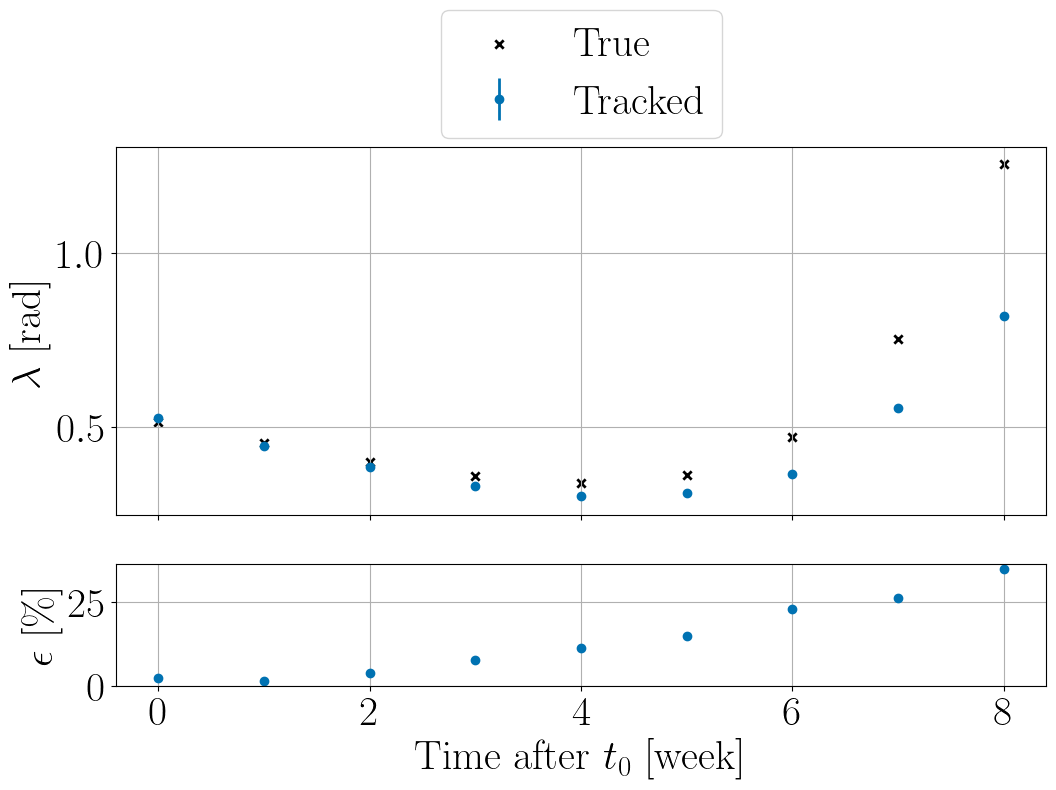}
        \caption{Source longitude $\lambda$ obtained from the analysis of $\tilde{\kappa}$ (tracked), known longitude (true) and absolute relative error ($\epsilon$) as a function of snapshot time.}
        \label{fig:async_lambda}
    \end{subfigure}
    \caption{Asynchronous case: latitude and longitude estimation as a function of snapshot time.}
    \label{fig:async_time}
\end{figure}

In addition, in Fig. \ref{fig:sync_time} and Fig. \ref{fig:async_time} we plot a comparison between expected and estimated values of latitude and longitude as a function of snapshot time. We also plot the absolute relative error between known and estimated values. In the synchronous case, Fig. \ref{fig:sync_time}, results show that an estimation of sky position based on $\tilde{\kappa}$ can be biased in this scenario and within a band of $\pm 0.3$ rad around the plane of the constellation $\tilde{\kappa}$ sky localization capabilities deteriorate. The biases induced by the motion of the constellation deserve to be studied in more detail in the future. In the asynchronous case, Fig. \ref{fig:async_time}, we focus on the first eight weeks after initial time $t_0$. The discrepancy between known and estimated $\beta$ and $\lambda$, captured by $\epsilon$, increases over time. As a matter of fact, after 4 weeks the mismatch between expected and estimated values reaches 5\% for latitude and 11\% for longitude. For comparison, after the same amount of time, $\epsilon$ reaches 1\% for latitude and 2\% for longitude in the synchronous case.

\begin{figure}[ht!]
    \begin{subfigure}[b]{0.48\textwidth}
        \centering
        \includegraphics[scale=0.28]{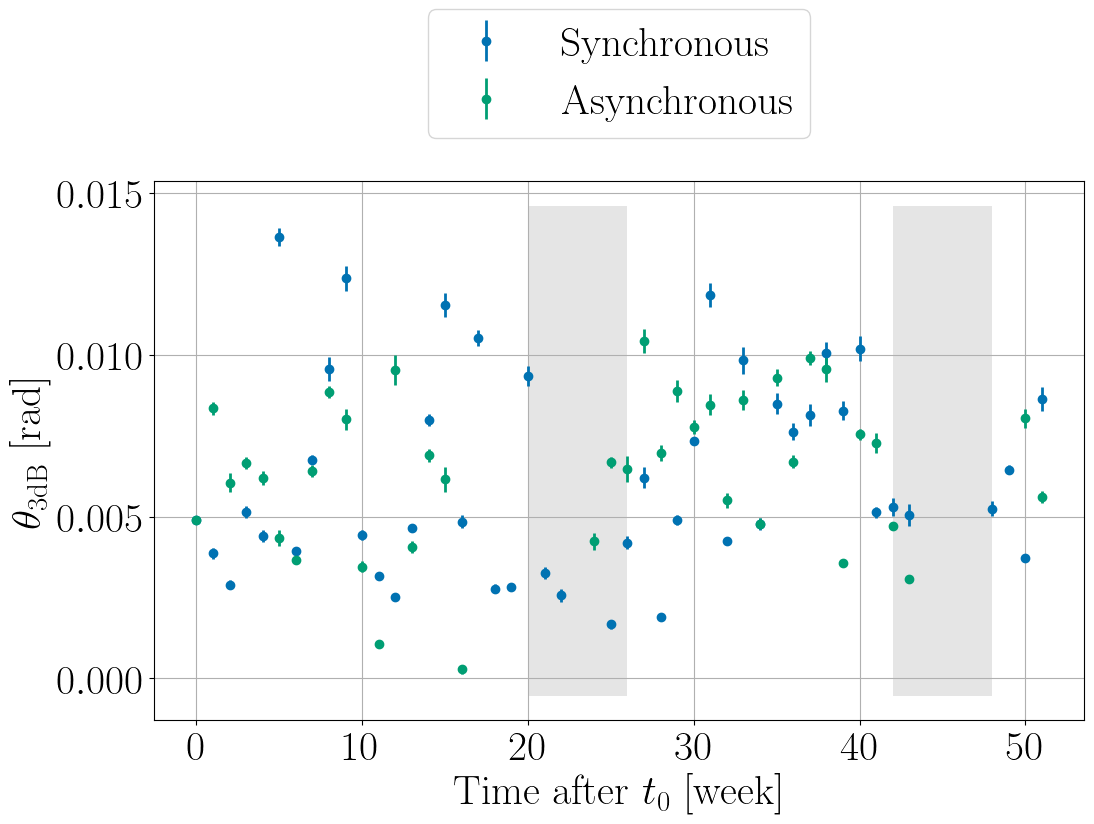}
        \caption{Angular resolution as a function of snapshot time. The gray area corresponds to the time when the apparent latitude of the source in the LISA frame is of $|\beta| < 0.3$ rad.}
        \label{fig:comp_resolution}
    \end{subfigure}
    \hfill
    \begin{subfigure}[b]{0.48\textwidth}
        \centering
        \includegraphics[scale=0.28]{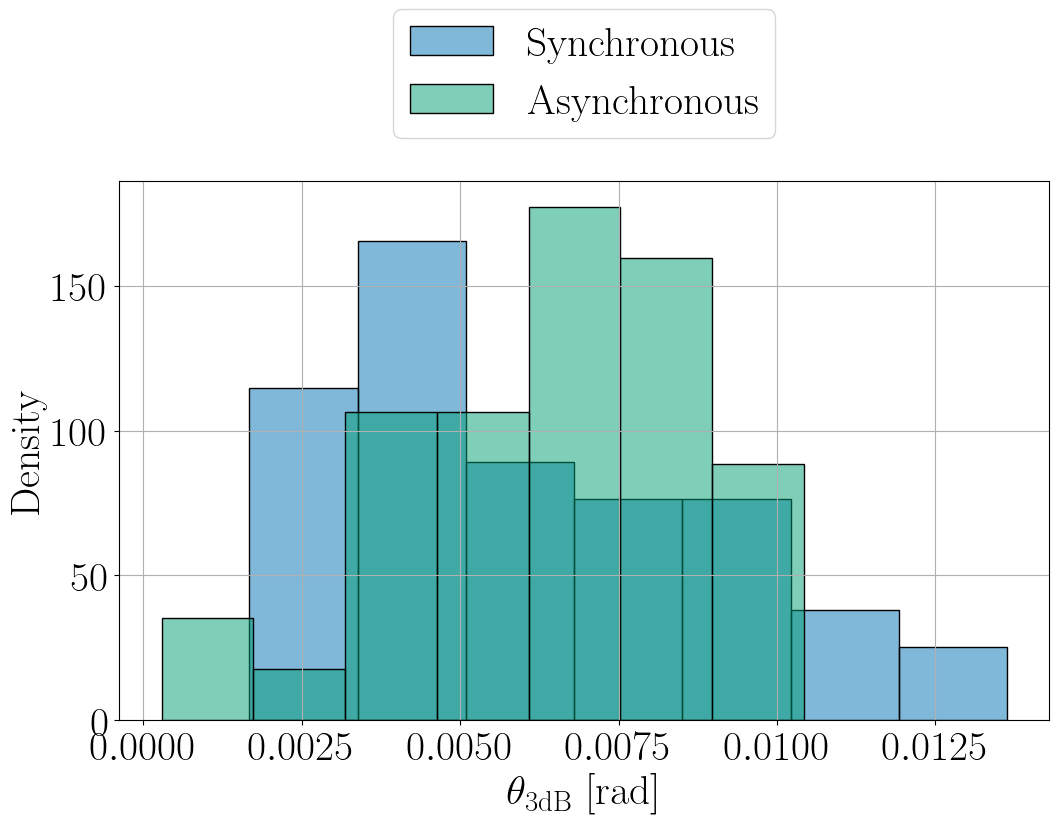}
        \caption{Angular resolution distribution. Histograms are normalized to one.}
        \label{fig:hist_resolution}
    \end{subfigure}
    \caption{Angular resolution as a function of snapshot time and angular resolution distribution.}
    \label{fig:resolution}
\end{figure}

Finally, Fig. \ref{fig:comp_resolution} shows the evolution of angular resolution as a function of snapshot time in both synchronous and asynchronous cases. In the synchronous case, the variation in angular resolution is compatible with the variations observed in Fig. \ref{fig:position_vgb}, i.e. the resolution evolves according to the apparent position of the source in the LISA frame. In the asynchronous case, a similar evolution is observed. In fact, Fig. \ref{fig:hist_resolution} displays the angular resolution distributions obtained in both cases. Distributions have compatible mean values at $(6.24 \pm 0.22) \cdot 10^{-3}$ rad for the synchronous case and at $(6.38 \pm 0.23) \cdot 10^{-3}$ rad for the asynchronous case. Thus, contrary to the estimated position of the source, angular resolution is not affected by the mismatch introduced in the asynchronous case.

\section{Conclusion}
The goal of this paper was to characterize a sky localization method based on $\tilde{\kappa}$ and therefore to set the ground for future applications of coronagraphic TDI. In particular, shifting away from purely theoretical tests of the technique, we have investigated $\tilde{\kappa}$'s response to both different types of signal and different assumptions on the LISA instrument.

First, we implemented a coronagraphic TDI calculation interfaced with state-of-the-art simulation software developed by the LISA Consortium. This way, we have tested the method on typical LISA sources, namely GBs and MBHBs, under simplifying assumptions. Through studying GBs we have found that the position of the source plays an important role on the angular resolution of the method. Importantly, for MBHBs we have shown that $\tilde{\kappa}$ can be used to estimate the sky position of such sources. Among tested MBHBs, whose parameters have been randomly drawn from an astrophysical catalog, the best angular resolution achieved was $(6.89 \pm 3.01) \, \mathrm{deg}^2$. 

Second, we have computed coronagraphic TDI's response to an analytic noise model, $S_{\kappa\kappa}$. We have found that the quantity $|\tilde{\kappa}|^2/S_{\kappa\kappa}$, even in the noiseless case, has can correct for the geometric property of $\tilde{\kappa}$ to attenuate signal near the plane of the constellation. Furthermore, we have examined coronagraphic TDI's response when considering realistic LISA orbits. We have found that in this scenario coronagraphic TDI's blind region around the plane of the constellation is enlarged. Moreover, when not updating coronagraphic TDI's coefficients according to the evolution of LISA orbits we have found a mismatch between the expected and estimated position of the source of 5\% on latitude and of 11\% on longitude after 4 weeks.

To conclude, coronagraphic TDI brings a promising model-agnostic approach to LISA data analysis. It makes a computationally efficient method which does not require template generation. Its efficiency stems both from the fact that coefficients entering the computation of $\tilde{\kappa}$ can be precomputed and that it focuses on a restricted parameter space. The results presented in this paper validate the applicability of coronagraphic TDI to the sky localization of typical LISA sources when considering a simplified LISA instrument. In the future, we aim at gauging coronagraphic TDI's sky localization performance in more realistic scenarios. In particular, within the context of low-latency searches, we are interested in further investigating $\tilde{\kappa}$'s ability to estimate the sky position of MBHBs. Finally, other potential applications of coronagraphic TDI, such as its use as a glitch veto, are also worth exploring in the future.

\begin{acknowledgments}
The authors thank J.-B. Bayle and A. Petiteau for fruitful discussions and useful remarks on improving this manuscript. The authors also thank the LISA Simulation Expert Group together with the LISA Data Challenge Working Group for the discussions on all simulation-related activities. Some of the results in this paper have been derived using the \textsc{healpy} and \textsc{HEALPix} packages.
\end{acknowledgments}

\appendix
\section{} \label{sec:app_definition}
Key to coronagraphic TDI is the decomposition given in Eq.\eqref{eq:decomposition}. As a matter of fact, vector $\vec{A}$ is computed from the functions $\alpha_{+, \times}$, $\beta_{+, \times}$ and $\gamma_{+, \times}$. In the following we derive $\alpha_{+, \times}$ from LISA's response to GWs and the definition of first generation TDI variable $\alpha$. The calculation presented here is analogous to \cite{PhysRevD.70.062002}, but is consistent with the notation used in \textsc{LISA GW Response}.

About LISA's response to GWs, we just collect the key formulas of the derivation of $y_{ij}$, the relative frequency shift of a photon traveling along link $ij$, provided in \textsc{LISA GW Response}'s documentation \cite{gwresponse}. Let us define the basis vectors
\begin{align}
    \hat{k} &= \begin{pmatrix}
        -\cos{\beta} \cos{\lambda} \\
        -\cos{\beta} \sin{\lambda} \\
        -\sin{\beta}
    \end{pmatrix} \label{eq:k} \\
    \hat{u} &= \begin{pmatrix}
        \sin{\lambda} \\
        - \cos{\lambda} \\
        0
    \end{pmatrix}\label{eq:u} \\
    \hat{v} &= \begin{pmatrix}
        - \sin{\beta} \cos{\lambda}\\
        -\sin{\beta}\sin{\lambda}\\
        \cos{\beta}
    \end{pmatrix} \label{eq:v}
\end{align}
where $\hat{k}$ describes the direction of propagation of the wave from the source to LISA and $\hat{u}$ and $\hat{v}$ are the reference polarization vectors. Keeping the indexing convention used in the main text, namely $j$ being the emitting spacecraft and $i$ being the receiving spacecraft, the deformation on link $ij$ induced by a GW with plus polarization $h_+$ and cross polarization $h_\times$ is given by 
\begin{equation}
    H_{ij}(t) = h_+(t) \xi^+(\hat{u}, \hat{v}, \hat{n}_{ij}) + h_\times(t) \xi^\times(\hat{u}, \hat{v}, \hat{n}_{ij})
\end{equation}
where $\xi^{+, \times}$ are the antenna patter functions and $\hat{n}_{ij}$ is the unit vector going from spacecraft $j$ to spacecraft $i$. Antenna pattern functions are defined by
\begin{align}
    \xi^+(\hat{u}, \hat{v}, \hat{n}_{ij}) &= (\hat{u} \cdot \hat{n}_{ij})^2 - (\hat{v} \cdot \hat{n}_{ij})^2 \\
    \xi^\times(\hat{u}, \hat{v}, \hat{n}_{ij}) &= 2(\hat{u} \cdot \hat{n}_{ij}) (\hat{v} \cdot \hat{n}_{ij})
\end{align}
where we assume $\hat{n}_{ij}$ does not depend on time. So in frequency domain we have
\begin{equation}
    \tilde{H}_{ij}(f) = \tilde{h}_+(f) \xi^+(\hat{u}, \hat{v}, \hat{n}_{ij}) + \tilde{h}_\times(f) \xi^\times(\hat{u}, \hat{v}, \hat{n}_{ij}).
\end{equation}

Moreover, the relative frequency shift of a photon traveling along link $ij$ is given by: 
\begin{widetext}
    \begin{equation}
        y_{ij}(t) \approx \frac{1}{2[1 - \hat{k} \cdot \hat{n}_{ij}(t)]} \bigg[H_{ij}\bigg(t - \frac{L_{ij}(t)}{c} - \frac{\hat{k}\cdot\vec{x}_j(t)}{c}\bigg) - H_{ij}\bigg(t - \frac{\hat{k}\cdot\vec{x}_i(t)}{c}\bigg)\bigg].
    \end{equation}
\end{widetext}
Assuming the constellation to be static, we can drop the time dependence of $\hat{n}_{ij}$, $L_{ij}$, $\vec{x}_{i}$ and $\vec{x}_{j}$, so the expression above simplifies to
\begin{equation}
    y_{ij}(t) = \frac{H_{ij}(t - L_{ij} - \hat{k}\cdot\vec{x}_j) - H_{ij}(t - \hat{k}\cdot\vec{x}_i)}{2(1 - \hat{k} \cdot \hat{n}_{ij})},
\end{equation}
where we set $c=1$ for convenience. Then in frequency domain we obtain
\begin{equation}
    \tilde{y}_{ij}(f) = \frac{e^{i\omega(L_{ij} + \hat{k}\cdot\vec{x}_j)} - e^{i\omega(\hat{k}\cdot\vec{x}_i)}}{2(1 - \hat{k} \cdot \hat{n}_{ij})} \tilde{H}_{ij}(f).
\end{equation}

Furthermore, the $y_{ij}$ enter the TDI algorithm through the definition of first generation Sagnac variable $\alpha$,
\begin{align}
    \alpha &= y_{12} + \bm{D}_{12}y_{23} + \bm{D}_{12}\bm{D}_{23}y_{31} \nonumber \\
    & \quad - (y_{13} + \bm{D}_{13}y_{32} + \bm{D}_{13}\bm{D}_{32}y_{21}),
\end{align}
where $\bm{D}_{ij}$ is the delay operator which delays any time series $x(t)$ by $L_{ij}$, that is $\bm{D}_{ij} x(t) = x(t-L_{ij})$. Through explicitly writing $\alpha$ in frequency domain we can define the coefficients
\begin{align}
    \alpha_{12} &= \frac{e^{i\omega(L_{12} + \hat{k}\cdot\vec{x}_2)} - e^{i\omega(\hat{k}\cdot\vec{x}_1)}}{2(1 - \hat{k} \cdot \hat{n}_{12})}\\
    \alpha_{23} &= \frac{e^{i\omega(L_{12} + L_{23} + \hat{k}\cdot\vec{x}_3)} - e^{i\omega(L_{12} + \hat{k}\cdot\vec{x}_2)}}{2(1 - \hat{k} \cdot \hat{n}_{23})}\\
    \alpha_{31} &= \frac{e^{i\omega(L_{12} + L_{23} + L_{31} + \hat{k}\cdot\vec{x}_1)} - e^{i\omega(L_{12} + L_{23} + \hat{k}\cdot\vec{x}_3)}}{2(1 - \hat{k} \cdot \hat{n}_{31})}\\
    \alpha_{13} &= \frac{e^{i\omega(\hat{k}\cdot\vec{x}_1)} - e^{i\omega(L_{13} + \hat{k}\cdot\vec{x}_3)}}{2(1 - \hat{k} \cdot \hat{n}_{13})}\\
    \alpha_{32} &= \frac{e^{i\omega(L_{13} + \hat{k}\cdot\vec{x}_3)} - e^{i\omega(L_{13} + L_{32} + \hat{k}\cdot\vec{x}_2)}}{2(1 - \hat{k} \cdot \hat{n}_{32})}\\
    \alpha_{21} &= \frac{e^{i\omega(L_{13} + L_{32} + \hat{k}\cdot\vec{x}_2)} - e^{i\omega(L_{13} + L_{32} + L_{21} + \hat{k}\cdot\vec{x}_1)}}{2(1 - \hat{k} \cdot \hat{n}_{21})}
\end{align}
such that if one defines  
\begin{equation}
    \alpha_{+,\times} = \sum_{ij} \alpha_{ij} \xi^{+,\times}(\hat{u}, \hat{v}, \hat{n}_{ij}),
\end{equation}
with $ij$ pairs running through $\{31, 13, 23, 32, 12, 21\}$, then one can write
\begin{equation}
    \tilde{\alpha}^{gw} = \alpha_+ \tilde{h}_+ + \alpha_\times \tilde{h}_\times.
\end{equation}
Note that in the main text we used the shorthand notation $\xi_{ij}^{+,\times} = \xi^{+,\times}(\hat{u}, \hat{v}, \hat{n}_{ij})$ for antenna pattern functions. Moreover, coefficients $\beta_{+, \times}$ and $\gamma_{+, \times}$ are derived from $\alpha_{+, \times}$ by cyclic permutation over the indices.

In paper \cite{PhysRevD.70.062002}, this derivation is done with a different notation. In particular, in \cite{PhysRevD.70.062002}, because orbits are assumed to be static and therefore $\hat{n}_{ij} = - \hat{n}_{ji}$, the calculation is done with just three LISA links. That is, instead of six $\alpha_{ij}$, just three $\alpha_k$ are obtained. In order to link the two it is necessary to map the $ij$-indexing to the $k$-indexing. This mapping is given by
\begin{equation}
    \hat{n}_{1} = \hat{n}_{23}, \quad \hat{n}_{2} = \hat{n}_{31}, \quad \hat{n}_{3} = \hat{n}_{12}.
\end{equation}
Then the $\alpha_i$ with $i \in \{1, 2, 3\}$ in paper \cite{PhysRevD.70.062002} correspond to:
\begin{align}
    - \alpha_1 &= \alpha_{32} + \alpha_{23} \\
    - \alpha_2 &= \alpha_{13} + \alpha_{31} \\
    - \alpha_3 &= \alpha_{21} + \alpha_{12}.
\end{align}

Finally, time domain coronagraphic TDI is simply the inverse Fourier transform of $\tilde{\kappa}$. Our definition differs from the one in \cite{PhysRevD.70.062002} just in a few points which are collected here. Using the same notation as in \cite{PhysRevD.70.062002}, we define $\kappa$ as
\begin{align}
    \kappa(t, \beta, \lambda) &= \sum_{k=1}^{27} [ A_1^{(k)}\alpha(t+\Delta_1^{(k)}) + A_2^{(k)}\beta(t+\Delta_2^{(k)}) \nonumber \\
    & \quad + A_3^{(k)}\gamma(t+\Delta_3^{(k)})],
\end{align}
and coefficients $\Delta_i^{(k)}$ in our implementation are the same as in \cite{PhysRevD.70.062002} with the exception of
\begin{equation}
    \Delta_3^{(13)} = L_1 + 2L_2 + l(\mu_1 + \mu_3), 
\end{equation}
from which $\Delta_1^{(13)}$ and $\Delta_2^{(13)}$ can be obtained by cyclic permutation over the indices. For a complete derivation, details on the notation used and the definition of $A_i^{(k)}$ the reader is referred to \cite{PhysRevD.70.062002}.

\section{} \label{sec:app_latitude}
The fact that $\tilde{\kappa}$ vanishes at zero latitude might be surprising. As mentioned in the text, this comes from the dependence of vector $\vec{A}$ on the antenna pattern functions. From the definitions given in App. \ref{sec:app_definition} we show that $\vec{A}$ is exactly zero at $\beta = 0$.

First, let us evaluate vector $\hat{v}$ given in Eq.\eqref{eq:v}. At $\beta=0$ and for any $\lambda$ we have
\begin{equation}
    \hat{v} = \begin{pmatrix}
        - \sin{\beta} \cos{\lambda}\\
        -\sin{\beta}\sin{\lambda}\\
        \cos{\beta}
    \end{pmatrix} = \begin{pmatrix}
        0\\
        0\\
        1
    \end{pmatrix}. 
\end{equation}
In addition, in the LISA frame all $\hat{n}_{ij}$ have a zero $z$-coordinate so the scalar product $\hat{v} \cdot \hat{n}_{ij}$ is zero. Hence,
\begin{equation}
    \xi^\times(\hat{u}, \hat{v}, \hat{n}_{ij}) = 2(\hat{u} \cdot \hat{n}_{ij}) (\hat{v} \cdot \hat{n}_{ij}) = 0.
\end{equation}

Now, the antenna pattern functions enter the coronagraphic TDI calculation via the functions $\alpha_{+, \times}$. In particular,
\begin{equation}
    \alpha_\times = \sum_{ij} \alpha_{ij} \xi^\times(\hat{u}, \hat{v}, \hat{n}_{ij}) = 0
\end{equation}
and similarly for $\beta_\times$ and $\gamma_\times$ which are also zero. Then following Eq. \eqref{eq:A} vector $\vec{A}$ reads
\begin{equation}
    \vec{A}(f, 0, \lambda) =  \begin{pmatrix}
        \beta_+ \gamma_\times - \gamma_+ \beta_\times \\
        \gamma_+ \alpha_\times - \alpha_+ \gamma_\times \\
        \alpha_+ \beta_\times - \beta_+ \alpha_\times
    \end{pmatrix} = \vec{0}.
\end{equation}
As a consequence Eq. \eqref{eq:def} writes
\begin{equation}
    \tilde{\kappa}(f, \beta, \lambda) = \vec{A}(f, 0, \lambda) \cdot \vec{D}(f) = 0
\end{equation}
for any vector $\vec{D}$. So $\tilde{\kappa}$ is zero at the plane of the constellation for both data and noise. 

\section{} \label{sec:app_signal}
\begin{figure}[ht!]
    \centering
    \includegraphics[scale=0.28]{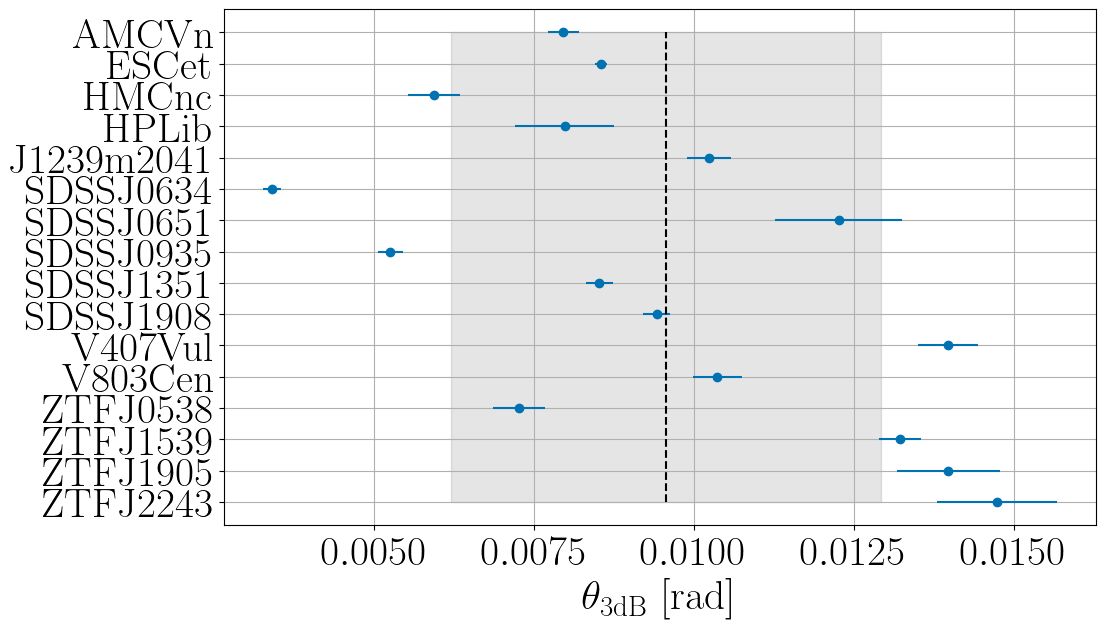}
    \caption{Angular resolutions and associated errors obtained for the 16 studied VGBs. The dashed line corresponds to the mean resolution and the gray area represents the standard deviation. The mean value is of $9.57 \cdot 10^{-3}$ rad, that is  $1.05$ deg$^2$.} \label{fig:app_vgb}
\end{figure}

Supplementary material relative to Sec. \ref{sec:signal} is gathered here. In Fig. \ref{fig:app_vgb}, we plot the angular resolution measured for all 16 analyzed VGBs. In Tab. \ref{tab:app_vgb} and Tab. \ref{tab:app_mbhb} we provide the mean and standard deviation of distributions in Fig. \ref{fig:violin_vgb} and Fig. \ref{fig:violin_mbhb} respectively. In Tab. \ref{tab:app_mbhb_params} we give a selection of parameters used to simulate studied MBHBs.

\begin{table}[ht!]
\caption{\label{tab:app_vgb}%
Mean ($\mu$) and standard deviation ($\sigma$) of distributions in Fig. \ref{fig:violin_vgb}.} 
\begin{ruledtabular}
\begin{tabular}{ccccc}
&\multicolumn{2}{c}{$\Omega_{3\mathrm{dB}} \, [\mathrm{deg}^2]$}&\multicolumn{2}{c}{$\Delta\Omega_{3\mathrm{dB}} \, [\mathrm{deg}^2]$}\\
Dataset & $\mu$ & $\sigma$ & $\mu$ & $\sigma$ \\ \hline
$\mathcal{A}$ & $9.15 \cdot 10^{-1}$ & $2.79 \cdot 10^{-9}$ & $3.99 \cdot 10^{-2}$ & $5.18 \cdot 10^{-10}$ \\
$f_0$ & $9.42 \cdot 10^{-1}$ & $2.12 \cdot 10^{-2}$ & $4.02 \cdot 10^{-2}$ & $1.51 \cdot 10^{-3}$ \\
$\dot{f}$ & $9.15 \cdot 10^{-1}$ & $3.99 \cdot 10^{-6}$ & $3.99 \cdot 10^{-2}$ & $8.30 \cdot 10^{-8}$ \\
$\iota$ & $9.49 \cdot 10^{-1}$ & $4.72 \cdot 10^{-2}$ & $6.33 \cdot 10^{-2}$ & $2.74 \cdot 10^{-2}$ \\
$\phi$ & $9.39 \cdot 10^{-1}$ & $1.79 \cdot 10^{-2}$ & $3.94 \cdot 10^{-2}$ & $4.26 \cdot 10^{-4}$ \\
$\Psi$ & $9.36 \cdot 10^{-1}$ &  $1.70 \cdot 10^{-2}$ & $3.95 \cdot 10^{-2}$ & $4.21 \cdot 10^{-4}$ \\
$(\beta_\star, \lambda_\star)$ & $1.28$ & $5.42 \cdot 10^{-1}$ & $1.43 \cdot 10^{-1}$ & $1.10 \cdot 10^{-1}$ \\
\end{tabular}
\end{ruledtabular}
\end{table}

\begin{table}[ht!]
\caption{\label{tab:app_mbhb}%
Mean ($\mu$) and standard deviation ($\sigma$) of distributions in Fig. \ref{fig:violin_mbhb}.} 
\begin{ruledtabular}
\begin{tabular}{ccccc}
&\multicolumn{2}{c}{$\Omega_{3\mathrm{dB}} \, [\mathrm{deg}^2]$}&\multicolumn{2}{c}{$\Delta\Omega_{3\mathrm{dB}} \, [\mathrm{deg}^2]$}\\
Dataset & $\mu$ & $\sigma$ & $\mu$ & $\sigma$ \\ \hline
MBHB\_00 & $15.0$ & $34.7$ & $14.5$ & $58.9$ \\
MBHB\_01 & $5.31$ & $20.6$ & $5.65$ & $34.7$ \\
MBHB\_02 & $22.5$ & $44.4$ & $14.3$ & $39.7$ \\
MBHB\_03 & $17.5$ & $38.5$ & $8.92$ & $25.9$ \\
MBHB\_04 & $9.60$ & $24.6$ & $5.97$ & $24.2$ \\
MBHB\_05 & $26.1$ & $53.0$ & $22.1$ & $71.0$ \\
MBHB\_06 & $22.9$ & $45.6$ & $20.1$ & $59.2$ \\
MBHB\_07 & $17.5$ & $42.4$ & $10.1$ & $33.8$ \\
MBHB\_08 & $22.9$ & $41.9$ & $18.3$ & $58.7$ \\
MBHB\_09 & $25.0$ & $52.2$ & $15.8$ & $41.8$ \\
MBHB\_10 & $6.51$ & $29.6$ & $3.68$ & $19.7$ \\
MBHB\_11 & $12.9$ & $26.7$ & $10.5$ & $44.5$ \\
MBHB\_12 & $78.1$ & $45.2$ & $116$ & $85.8$ \\
MBHB\_13 & $22.0$ & $44.8$ & $17.8$ & $50.2$ \\
MBHB\_14 & $31.9$ & $44.8$ & $31.2$ & $68.8$ \\
\end{tabular}
\end{ruledtabular}
\end{table}

\begin{table*}[ht!]
\caption{\label{tab:app_mbhb_params}%
Selected simulation parameters for MBHBs.
}
\begin{ruledtabular}
\begin{tabular}{ccccccc}
\textrm{Source}&
\textrm{$\beta_\star$ [rad]}&
\textrm{$\lambda_\star$ [rad]}&
\textrm{$M_1$ [$M_\odot$]}&
\textrm{$M_2$ [$M_\odot$]}&
\textrm{$M$ [$M_\odot$]}&
\textrm{$z$}\\
\colrule
    MBHB\_00 & $-5.64 \cdot 10^{-1}$ & $6.11 \cdot 10^{-1}$ & $1.02 \cdot 10^{6}$ & $7.97 \cdot 10^{5}$ & $1.81 \cdot 10^{6}$ & 2.18\\
    MBHB\_01 & -1.08 & 4.05 & $4.96 \cdot 10^{6}$ & $4.07 \cdot 10^{6}$ & $9.02 \cdot 10^{6}$ & 6.18\\
    MBHB\_02 & $6.36 \cdot 10^{-1}$ & $7.79 \cdot 10^{-1}$ & $1.77 \cdot 10^{6}$ & $1.08 \cdot 10^{6}$ & $2.86 \cdot 10^{6}$ & 6.03\\
    MBHB\_03 & $-8.79 \cdot 10^{-1}$ & 4.55 & $9.16 \cdot 10^{5}$ & $7.02 \cdot 10^{5}$ & $1.62 \cdot 10^{6}$ & 1.10\\
    MBHB\_04 & $-3.03 \cdot 10^{-1}$ & 1.29 & $1.32 \cdot 10^{6}$ & $6.12 \cdot 10^{5}$ & $1.94 \cdot 10^{6}$ & 1.74\\
    MBHB\_05 & $1.47 \cdot 10^{-1}$ & 5.99 & $2.83 \cdot 10^{6}$ & $2.32 \cdot 10^{6}$ & $5.15 \cdot 10^{6}$ & 5.51\\
    MBHB\_06 & $-9.87 \cdot 10^{-1}$ & 3.89 & $3.20 \cdot 10^{6}$ & $3.06 \cdot 10^{6}$ & $6.26 \cdot 10^{6}$ & 2.88\\
    MBHB\_07 & $9.17 \cdot 10^{-1}$ & 3.58 & $1.86 \cdot 10^{6}$ & $1.80 \cdot 10^{6}$ & $3.66 \cdot 10^{6}$ & 4.57\\
    MBHB\_08 & $8.35 \cdot 10^{-1}$ & 3.26 & $3.98 \cdot 10^{6}$ & $2.05 \cdot 10^{6}$ & $6.03 \cdot 10^{6}$ & 3.35\\
    MBHB\_09 & $-5.21 \cdot 10^{-1}$ & 3.09 & $3.13 \cdot 10^{6}$ & $2.64 \cdot 10^{6}$ & $5.77 \cdot 10^{6}$ & 7.21\\
    MBHB\_10 & 1.29 & 2.11 & $1.35 \cdot 10^{6}$ & $1.20 \cdot 10^{6}$ & $2.54 \cdot 10^{6}$ & 3.59\\
    MBHB\_11 & $9.48 \cdot 10^{-2}$ & 5.18 & $1.68 \cdot 10^{6}$ & $6.08 \cdot 10^{5}$ & $2.28 \cdot 10^{6}$ & 4.31\\
    MBHB\_12 & $-4.41 \cdot 10^{-2}$ & 3.78 & $3.58 \cdot 10^{6}$ & $1.49 \cdot 10^{6}$ & $5.06 \cdot 10^{6}$ & 7.19\\
    MBHB\_13 & $-3.17 \cdot 10^{-1}$ & 1.30 & $1.24 \cdot 10^{6}$ & $1.14 \cdot 10^{6}$ & $2.37 \cdot 10^{6}$ & 4.45\\
    MBHB\_14 & $1.03 \cdot 10^{-1}$ & $2.24 \cdot 10^{-1}$ & $6.39 \cdot 10^{6}$ & $3.16 \cdot 10^{6}$ & $9.55 \cdot 10^{6}$ & 8.08\\
\end{tabular}
\end{ruledtabular}
\end{table*}

\section{} \label{sec:app_rotation}
The link between the LISA and SSB frame is done thanks to a rotation matrix which we derive here. Let $\vec{x}_i$ be the position vector of spacecraft $i$ in the SSB frame, with $i \in \{1, 2, 3\}$. Then, introducing the position of the constellation's barycenter $\vec{x}_0$, we define the vectors
\begin{equation}
    \hat{q}_i = \frac{\vec{x}_i - \vec{x}_0}{||\vec{x}_i - \vec{x}_0||}.
\end{equation}
From the $\hat{q}_i$ vectors we construct the basis vectors $\{\hat{x}_L, \hat{y}_L, \hat{z}_L\}$ which define the LISA frame:
\begin{equation}
    \hat{x}_L = \hat{q}_1, \quad \hat{z}_L = \frac{\hat{q}_1 \times \hat{q}_3}{||\hat{q}_1 \times \hat{q}_3||}, \quad \hat{y}_L = \hat{z}_L \times \hat{x}_L.
\end{equation}
The the rotation matrix $R$ which transforms a vector $\vec{x}$ in the SSB frame into a vector $\vec{x'}$ in the LISA frame is given by
\begin{equation}
    R = (\hat{x}_L \, \hat{y}_L \, \hat{z}_L)^{-1}.
\end{equation}

Now consider some GW source and let $\hat{k}_B$ be the vector which points from this GW source to the origin of the SSB frame and $\hat{k}_L$ be the vector which points from the same source to the origin of the LISA frame. We define
\begin{equation}
    \hat{k}_B = \begin{pmatrix}
        -\cos{\beta_B} \cos{\lambda_B} \\
        -\cos{\beta_B} \sin{\lambda_B} \\
        -\sin{\beta_B}
    \end{pmatrix} \,
    \hat{k}_L = \begin{pmatrix}
        -\cos{\beta_L} \cos{\lambda_L} \\
        -\cos{\beta_L} \sin{\lambda_L} \\
        -\sin{\beta_L}
    \end{pmatrix}.
\end{equation}
Then $(\beta_B, \lambda_B)$ is the position of the source in the SSB frame and $(\beta_L, \lambda_L)$ is the position in the LISA frame. The two vectors are related through the rotation $R$:
\begin{equation}
    R \cdot \hat{k}_B = \hat{k}_L. \label{eq:frame}
\end{equation}
Noting $R_{ij}$ the matrix elements of $R$ and working out Eq. \eqref{eq:frame}, we obtain the following relation between $(\beta_B, \lambda_B)$ and $(\beta_L, \lambda_L)$  
\begin{widetext}
    \begin{align}
    \beta_L &= \arcsin[(R_{20} \cos{\lambda_B} + R_{21}\sin{\lambda_B})\cos{\beta_B} + R_{22}\sin{\beta_B}] \\
    \lambda_L &= \arctan\bigg[\frac{(R_{10} \cos{\lambda_B} + R_{11}\sin{\lambda_B})\cos{\beta_B} + R_{12}\sin{\beta_B}}{(R_{00} \cos{\lambda_B} + R_{01}\sin{\lambda_B})\cos{\beta_B} + R_{02}\sin{\beta_B}}\bigg]. \label{eq:bcrs_lisa}
    \end{align}  
\end{widetext}
Similarly, one can express $(\beta_B, \lambda_B)$ as a function of $(\beta_L, \lambda_L)$ using the inverse transformation $R^{-1}$.

\bibliography{references}

\end{document}